\documentclass[showpacs,preprintnumbers]{revtex4}
\usepackage{latexsym}

\usepackage[T1]{fontenc}

\usepackage[latin1]{inputenc}

\usepackage{graphicx}
\usepackage{color,pst-plot}

\usepackage{amsmath}
\usepackage {amsfonts,amssymb,amstext}

\newcommand{\lbl}[1]{\label{#1}}
\newcommand{ \rf}[1]{(\ref{#1})}

\newcommand{\be}{\begin{equation}}
\newcommand{\ee}{\end{equation}}{
\newcommand{\bea}{\begin{eqnarray}}
\newcommand{\eea}{\end{eqnarray}}
\newcommand{\setl}{\setlength\arraycolsep{2pt}}

\newcommand{\noi}{\noindent}
\newcommand{\nn}{\nonumber}
\newcommand{\ra}{\rightarrow}
\newcommand{\Ra}{\Rightarrow}

\newcommand{\cM}{{\cal M}}

\newcommand{\cO}{{\cal O}}

\newcommand{\cS}{{\cal S}}

\newcommand{\Imm}{\mbox{\rm Im}}
\newcommand{\Ree}{\mbox{\rm Re}}

\newcommand{\MeV}{\mbox{\rm MeV}}

\newcommand{\rH}{\mbox{\rm H}}

\newcommand{\annd}{\mbox{\rm and}}

\newcommand{\g}{\mbox{\bf g}}

\DeclareMathAlphabet{\eusm}{U}{}{}{}
\SetMathAlphabet\eusm{normal}{U}{eus}{m}{n}
\SetMathAlphabet\eusm{bold}{U}{eus}{b}{n}
\DeclareMathAlphabet{\mathpzc}{OT1}{pzc}{m}{it}

\input epsf
\setcounter{section}{0}

\setcounter{equation}{0}
\def\theequation{\arabic{section}.\arabic{equation}}

\begin{document}

%\preprint{CPT-P010-2008}
%\preprint{IFIC/08-06}

\title{ Muon Anomaly from Lepton Vacuum Polarization \\ and  the Mellin--Barnes Representation}

\author{Jean-Philippe Aguilar} 
\email{aguilar@cpt.univ-mrs.fr}
\affiliation{Centre  de Physique Th\'eorique CNRS-Luminy Case 907 F-13288 Marseille Cedex 9, France}
\author{David Greynat} 
\email{david.greynat@ific.uv.es}
\affiliation{IFIC, Universitat de Val\`encia-CSIC Apt. Correus 22085, E-46071 Val\`encia, Spain}
\author{Eduardo de Rafael}
\email{EdeR@cpt.univ-mrs.fr}
\affiliation{Centre  de Physique Th\'eorique CNRS-Luminy Case 907 F-13288 Marseille Cedex 9, France}

\begin{abstract}
We evaluate, analytically, a specific class of eighth--order and tenth--order QED contributions to the anomalous magnetic moment of the muon. They are generated by Feynman  diagrams involving lowest order vacuum polarization insertions of leptons $l=e,\mu$, and $\tau$. The results are given in the form of analytic expansions in terms of the mass ratios $m_e/m_\mu$ and $m_\mu/m_\tau$.  We compute as many terms as required by the error induced by the present experimental uncertainty on the lepton masses.  
We show how the Mellin--Barnes integral representation of Feynman parametric integrals allows for an easy analytic evaluation of as many terms as wanted in these expansions and how its underlying algebraic structure generalizes the standard renormalization group properties. We also discuss the generalization of this technique  to the case where two independent mass ratios appear. Comparison with previous numerical and analytic evaluations made in the literature, whenever pertinent, are also made.    
\end{abstract}

\maketitle

%%%%%%%%%%%%%%%%%%%%%%%%%%%%%%%%%%
\section{\small Introduction}\lbl{int}
\setcounter{equation}{0}
\def\theequation{\arabic{section}.\arabic{equation}}

The present experimental world average of the anomalous magnetic moment of the muon $a_{\mu}$, assuming CPT--invariance, viz. $a_{\mu^+} = a_{\mu^-}$,  is
\be\lbl{amuexp}
a_{\mu}^{(\rm exp)}=116~592~080~(63)\times 10^{-11}\quad (0.54~{\rm ppm})\,,
\ee
where the total uncertainty includes a 0.46 ppm statistical uncertainty and a 0.28 ppm systematic uncertainty, combined in quadrature.
This result is largely dominated by a series of precise measurements carried out  at the Brookhaven National Laboratory (BNL) during the last few years, by the E821 collaboration, with results reported in ref.~\cite{Bennet06} and references therein. The prediction of the Standard Model, as a result of contributions from many physicists is~\footnote{See e.g. the review article in ref.~\cite{MdeRR07} and references therein.}
\be\lbl{amuth}
a_{\mu}^{(\rm SM)}=116~591~785~(61)\times 10^{-11}\,,
\ee
where the error here is dominated at present  by the lowest order hadronic vacuum polarization  contribution uncertainty ($\pm 46.6\times 10^{-11}$), as well as by the theoretical uncertainty in the hadronic light--by--light scattering contribution, estimated to be $\pm 40\times 10^{-11}$. Errors here have also been combined in quadrature. The results quoted in~\rf{amuexp} and \rf{amuth} imply a 3.4 standard deviation in the difference
\be
a_{\mu}^{(\rm exp)}-a_{\mu}^{(\rm SM)}=(295\pm 88)\times 10^{-11}\,.
\ee

This $3.4~\sigma$ deviation deserves attention. Ideally, and before one can attribute the present discrepancy to new Physics,  one would like to reduce the theoretical uncertainties as much as possible, parallel to a new experimental effort toward an even more precise measurement of $a_{\mu}$~\cite{whitepaper,F04}. It is also important to reexamine critically  the various theoretical contributions to Eq.~\rf{amuth}; primarily the hadronic contributions of course, but also the higher order QED and electroweak  contributions. It would be reassuring to have, at least, two independent calculations of some of these contributions, as well as of the higher order estimates. The purpose of this article is a first step ({\it albeit a small one}) in this direction.

We shall be concerned with a specific class  of eighth--order and tenth--order QED contributions to $a_{\mu}$ illustrated by the Feynman diagrams shown in Fig.~1 and Fig.~2 below. The contribution from the diagrams in these figures with no $\tau$ vacuum polarization insertions i.e., diagrams (A), (B) and (C) in Fig.~1 and diagrams (A), (B), (C)  and (D) in Fig.~2, have  been estimated numerically by Kinoshita and Nio in refs.~\cite{KN04} and \cite{KN06} (see also ref.~\cite{NAHK07}). There are also analytic results of the same diagrams, published by Laporta~\cite{La93,La94}, in the form of analytic expansions in terms of the electron to muon  mass ratio $\frac{m_e}{m_\mu}$. To our knowledge, the diagrams in Fig.~1 (D) and Fig.~2 (E) involving a tau loop vacuum polarization insertion have not been considered before. We find for these contributions the following results:
\be
a_{\mu}^{(ee\tau)} =\left(\frac{\alpha}{\pi} \right)^4 \ 0.002~748~6(9)
\qquad\annd\qquad a_{\mu}^{(eee\tau)} =\left(\frac{\alpha}{\pi} \right)^5 0.013~057~4(4)\,,
\ee
where the errors are the ones induced by the present accuracy in the determination of the lepton masses. 
As one can see from the other results reported in the Appendix, the contribution from the diagrams in Fig.~2 (E) is of the same size as the one from the diagrams in Fig.~2 (D).

In a previous paper~\cite{FGdeR05}, it has been shown how the Mellin--Barnes integral representation of Feynman parametric integrals combined with the converse mapping theorem~\cite{FGD95}, allow for an easy evaluation of as many terms as wanted in the asymptotic behaviour of Feynman diagrams in terms of a mass ratio. We shall show how this technique applies to the evaluation of the class of diagrams considered here, and how to generalize it to the case where two independent mass ratios, like  $\frac{m_e}{m_\mu}$ and $\frac{m_\mu}{m_\tau}$ in the diagrams in Fig.~1 (D) and Fig.~2 (E), enter into consideration. We shall do that in a rather detailed way for various reasons: 
\begin{itemize}
	\item 
 We find, regretfully, that many analytic calculations found in the literature provide very few details about the techniques employed; yet it would be useful to compare the efficiency of different methods when planning new more complex calculations.
 \item Giving  details on some of the intermediate steps of complex calculations can only help to see the qualities and limitations of the methods employed and, occasionally, to spot potential errors.
 \item As explained in ref.~\cite{LdeR74} many years ago, there is a large class of higher order contributions which can be estimated using renormalization group arguments: {\it powers of $\log\frac{m_\mu}{m_e}$ terms at a given order are algebraically related to lower order contributions}. Subsequent applications have been made by many other authors~\footnote{For a recent application to $\cO(\alpha^4)$ and $\cO(\alpha^5)$  in QED see e.g. ref~\cite{Ka92,Ka06} and references therein.}; however, very little is known about how to extend  renormalization group arguments to subleading terms: {\it constant terms and powers of $\log\frac{m_\mu}{m_e}$ suppressed by $\frac{m_e}{m_\mu}$ powers}. The Mellin--Barnes technique that we are advocating provides a precise answer to that question: {\it as we shall see, in the Mellin--Barnes representation, the  asymptotic contributions at a given order factorize in terms of well defined moments of lower order contributions}. 
 \item
 Having a powerful technique to obtain asymptotic expansions provides a useful  alternative to the computation of exact analytic expressions, which are often very complicated and cumbersome. After all, ratios of masses are known from experiment only to a fixed accuracy; the associated  error propagates into a numerical uncertainty of the exact analytic result in any case. Computing as many terms in the corresponding asymptotic expansion, as required by the experimental precision in the masses involved, provides an easier alternative to the computation of an exact analytic result, {\it with the same practical accuracy}. The calculations in this paper have all been done  within this spirit.
 \item Finally, we think that it is important to have an independent way to check the precision of the contributions evaluated numerically. Many of the multidimensional integrals involved in higher order calculations are far from trivial, which has obliged the experts to develop skillful methods.  The numerical results are often dominated by a statistical error which is larger than the error induced by the experimental determination of the lepton masses. As we shall see, within this limited precision, we find that all the results of Kinoshita and collaborators~\cite{KN04,KN06,NAHK07} checked in this paper are correct, {\it  within less than one standard deviation of their estimated error}, which is a remarkable performance.
\end{itemize}

The paper has been organized as follows. The next section summarizes well known results about lepton vacuum polarization and the way we choose to express the various contributions to the muon anomalous magnetic moment. Section III gives a detailed description of the Mellin--Barnes technique and the converse mapping theorem as applied to the eighth--order and tenth--order contributions that we are considering. We also discuss in this section the underlying algebraic properties which generalize the usual renormalization group constraints previously discussed in the literature. Section IV contains a discussion of the evaluation of the various moment integrals which appear in the intermediate steps. They can all be done analytically and we give explicit expression for all of them; but we also show how in practice, the evaluation of a finite number of terms in the ultimate expansion, only requires the explicit knowledge of approximate expressions for these moments. This is important to know, in view of more complicated situations that one may very well encounter when considering other Feynman diagrams. 
The final evaluation of the eighth--order diagrams is done in Section V and of  the tenth--order diagrams in Section VI. Section VII is dedicated to the generalization of the techniques discussed in previous sections to the case where three mass scales appear. The calculation of the diagrams in Figs. 1 (D) and 2 (E) are non--trivial examples of applications of this new technology. Finally, we collect in an Appendix all the results that we have computed in this paper. 
    
%%%%%%%%%%%%%%%%%%%%%%%%%%%%%%%%%%%%%%%%%%%%%%%%%%%%%%%%%%%%%%%%%%%%%%%%
%%%%%%%%%%%%%%%%%%%%%%%%%%%%%%%%%%%%%%%%%%%%%%%%%%%%%%%%%%%%%%%%%%%%%%%%
\section{\small Photon Propagator and  the Muon Anomaly}\lbl{BSC}
\setcounter{equation}{0}
\def\theequation{\arabic{section}.\arabic{equation}}

We are interested in the replacement of a free photon--propagator by the dressed propagator
\be\lbl{fullpp}
D_{\alpha\beta}(q)=-i\left(g_{\alpha\beta}-\frac{q_{\alpha}q_{\beta}}{q^2}\right)\frac{1}{q^2} \frac{1}{1+\sum_{l}\Pi^{(l)}(q^2)}-ia\frac{q_{\alpha}q_{\beta}}{q^4}\,,
\ee
where $\Pi^{(l)}(q^2)$ denotes the  proper vacuum polarization self--energy  contribution induced by a lepton loop $l=e,\mu,\tau$ and $a$ is a parameter reflecting the gauge 
freedom in the free--field propagator ($a=1$ in the Feynman gauge). In fact, all the diagrams which we shall be considering here are gauge independent and, therefore, the term $i(1-a)\frac{q_{\alpha}q_{\beta}}{q^4}$ does not contribute to their evaluation.

 The perturbation theory expansion generates a  series in powers of the functions  $\Pi^{(l)}(q^2)$. Here we shall be concerned with a particular selection of the terms in this series, namely those with three and four powers of the lowest order $\Pi^{(l)}(q^2)$ self--energies; more precisely with the terms:
 
 {\setl
\bea
 \frac{1}{1+\sum_{l}\Pi^{(l)}(q^2)} & \doteq &  -\left[\Pi^{(e)}(q^2)\right]^3 -{\bf 3}\ \left[\Pi^{(e)}(q^2)\right]^2 \Pi^{(\mu)}(q^2) -{\bf 3}\ \Pi^{(e)}(q^2)  \left[\Pi^{(\mu)}(q^2)\right]^2 \lbl{3l}\\
 & & \ -{\bf 3}\ \left[\Pi^{(e)}(q^2)\right]^2 \Pi^{(\tau)}(q^2) \lbl{3tau}\\
 & &\hspace*{-4cm} + \left[\Pi^{(e)}(q^2)\right]^4 +{\bf 4}\ \left[\Pi^{(e)}(q^2)\right]^3 \Pi^{(\mu)}(q^2) +{\bf 6}\ \left[\Pi^{(e)}(q^2)\right]^2 \left[\Pi^{(\mu)}(q^2)\right]^2
  +{\bf 4}\  \Pi^{(e)}(q^2) \left[\Pi^{(\mu)}(q^2)\right]^3 \lbl{4l} \\
  & & \ +{\bf 4}\ \left[\Pi^{(e)}(q^2)\right]^3 \Pi^{(\tau)}(q^2)  \lbl{4tau}\,.
\eea}

\noi
When inserted in the free--photon propagator of the lowest order one loop muon vertex, these terms generate the four types of Feynman diagrams  collected in Fig.~1 and the five types of diagrams collected in Fig.~2. The Feynman diagrams in these figures give, respectively, {\it eighth order} and {\it tenth order}  contributions  to the anomalous magnetic moment of the muon, which are enhanced by powers of $\log\frac{m_{\mu}}{m_e}$ factors. This is why we are interested in these terms. The reason why we only keep the terms in \rf{3tau} and \rf{4tau} with  one power of $\Pi^{(\tau)}(q^2)$ is because, in spite of the enhancement by $\log\frac{m_{\mu}}{m_e}$ factors, the $\tau$--loop induces a suppression factor $\sim\frac{m_{\mu}^2}{m_{\tau}^2}$. Contributions  from two or more powers of $\Pi^{(\tau)}(q^2)$ factors to the muon anomaly will be, therefore, neglected. 

A very convenient integral representation for the contribution to the muon anomaly from the graphs in Fig.~1 and Fig.~2 can be obtained as follows. First we write a dispersion relation for the on--shell renormalized photon propagator induced by electron self--energy loops only. This generates the following set of relations:
\begin{equation}
\left[\Pi^{(e)}(q^2)\right]^j  =  \int_0^\infty\frac{d\mathsf{t}}{\mathsf{t}}\frac{q^2}{\mathsf{t}-q^2 -i\epsilon}\ \rho_{j}\left(\frac{4m_{e}^2}{\mathsf{t}}\right)\;,\qquad j=1\,,2\,,3\,,4\,,	
\end{equation}
with

{\setl
\bea
\rho_{1}\left(\frac{4m_{e}^2}{\mathsf{t}}\right) & = & \frac{1}{\pi}\Imm\Pi^{(e)}(\mathsf{t})\,, \lbl{dr1}\\
\rho_{2}\left(\frac{4m_{e}^2}{\mathsf{t}}\right) & = & \frac{1}{\pi}\left\{2\ \Ree\Pi^{(e)}(\mathsf{t})\ \Imm\Pi^{(e)}(\mathsf{t})\right\}\,, \lbl{dr2}\\
\rho_{3}\left(\frac{4m_{e}^2}{\mathsf{t}}\right) & = & \frac{1}{\pi}\left\{3\! \left[\Ree\Pi^{(e)}(\mathsf{t})\right]^2\Imm\Pi^{(e)}(\mathsf{t})-\left[\Imm\Pi^{(e)}(t)\right]^3\right\}\,, \lbl{dr3}\\
\rho_{4}\left(\frac{4m_{e}^2}{\mathsf{t}}\right)& = & \frac{1}{\pi}\left\{4\! \left[\Ree\Pi^{(e)}(\mathsf{t})\right]^3\!\Imm\Pi^{(e)}(\mathsf{t})-4\Ree\Pi^{(e)}(\mathsf{t})\!\left[\Imm\Pi^{(e)}(\mathsf{t})\right]^3\right\}\,. \lbl{dr4}
\eea}

\noi
Using these representations, the contribution to the muon anomaly from the various terms in the series in Eqs.~\rf{3l},~\rf{3tau} and \rf{4l},~\rf{4tau} can be viewed as the convolution of electron spectral functions, modulated by powers of the muon or tau self-energy functions, with a free--photon propagator replaced by a {\it fictitious massive photon} propagator:  
\begin{equation}
-ig_{\alpha\beta}\ \frac{1}{q^2}\Ra	-ig_{\alpha\beta}\ \frac{(-1)}{q^2 -\mathsf{t}+i\epsilon}\,.
\end{equation}
The muon vertex loop integral over the virtual $q$--momenta can then be traded by a Feynman $x$--parameter integral (see the review article in ref.~\cite{LPdeR72} and the earlier references therein), with a net contribution to the muon anomaly, up to an overall combinatorial factor $F_{(j,p)}$ which can be read from the expansion in Eqs.~\rf{3l} to \rf{4tau}:
\be\lbl{rep}
 a_{\mu}^{(j,p)}=	 \frac{\alpha}{\pi}\int_{0}^{\infty}\frac{d\mathsf{t}}{\mathsf{t}}
\int_0^1 dx\frac{x^2 (1-x)}{x^2 +\frac{\mathsf{t}}{m_{\mu}^2}(1-x)}\; F_{(j,p)}\;  	\left[\Pi^{(l=\mu\,,\tau)}\left(\frac{-x^2}{1-x}m_{\mu}^2\right)\right]^p \rho_{j}\left(\frac{4m_e^2}{\mathsf{t}}\right)\,, 
\ee
where $p=0,1,2,3$ when $l=\mu$ and $p=1$ when $l=\tau$, while the index $j$ counts the number of electron loops in the vacuum polarization; therefore
\be
F_{(3,0)}=1\,,\quad F_{(2,1)}=3\,,\quad F_{(1,2)}=3\,,
\ee
and
\be
F_{(4,0)}=-1\,,\quad F_{(3,1)}=-4\,,\quad F_{(2,2)}=-6\,,\quad F_{(1,3)}=-4\,.
\ee
Notice that in the process of trading the integral over the virtual $q$--momenta by the Feynman parameter $x$, one has obtained the effective replacement:
\begin{equation}\lbl{Q2x}
	q^2 \Ra \frac{-x^2}{1-x}m_{\mu}^2\,,
\end{equation}
 in the muon and tau self--energy functions.
 
 The lowest order vacuum polarization self--energy functions in QED are well known. We shall use the representations given in ref.~\cite{LdeR68}. With
\begin{equation}\lbl{delta}
	\delta=\sqrt{1-\frac{4m_{e}^2}{\mathsf{t}}}\,,
\end{equation}
 the lowest order spectral function for the electron is
\begin{equation}\lbl{losf}
 \frac{1}{\pi}\Imm\Pi^{(e)}(\mathsf{t}) = 	\frac{\alpha}{\pi}\delta\left(\frac{1}{2}-\frac{1}{6}\delta^2\right)\;\;\theta\left(\mathsf{t}-4m_{e}^2\right)\,,
\end{equation}
and the real part $\left(\delta=\sqrt{1-\frac{4m_{e}^2}{q^2}}\ \right)$ is
\begin{equation}\lbl{ree}
\Ree\Pi^{(e)}(q^2)=	\left(\frac{\alpha}{\pi}\right)
\left[\frac{8}{9}-\frac{1}{3}\delta^2+\delta\left(\frac{1}{2}-\frac{1}{6}\delta^2\right) 
\log\frac{\vert 1-\delta\vert }{1+\delta}\right]\,.
\end{equation}
We shall also need the expression for the vacuum polarization self--energy induced by a muon loop, in the euclidean region,  and as a function of the Feynman $x$--parameter:
\begin{equation}
\Pi^{(\mu)}\left(\frac{-x^2}{1-x}m_{\mu}^2\right)=	\left(\frac{\alpha}{\pi}\right)
\left[\frac{5}{9}+\frac{4}{3x}-\frac{4}{3x^2}+ \left(-\frac{1}{3}+\frac{2}{x^2}-\frac{4}{3x^3}\right)\log(1-x)\right]\,;
\end{equation}
as well as the one induced by a tau loop which, for our purposes, it is more convenient to keep in the Feynman parametric integral representation:
\begin{equation}\lbl{rtau}
\Pi^{(\tau)}\left(\frac{-x^2}{1-x}m_{\mu}^2\right)=-	\frac{\alpha}{\pi}
\int_0^1 dz 2z(1-z)\log\left[1+\frac{x^2}{1-x}z(1-z)\frac{m_{\mu}^2}{m_{\tau}^2} \right]\,.
\end{equation}

%%%%%%%%%%%%%%%%%%%%%%%%%%%%%%%%%%%%%%%%%%%%%%%%%%%%%%%%%%%%%%%%%%%%%%%%
%%%%%%%%%%%%%%%%%%%%%%%%%%%%%%%%%%%%%%%%%%%%%%%%%%%%%%%%%%%%%%%%%%%%%%%%
\section{\small The Mellin--Barnes Representation and the Renormalization Group}\lbl{MBR}
\setcounter{equation}{0}
\def\theequation{\arabic{section}.\arabic{equation}}

It is now useful to introduce the two Mellin--Barnes integral  representations~\footnote{See e.g. ref.~\cite{FGD95} and references therein. To our knowledge, the use of the Mellin--Barnes representation in Quantum Field Theory was first proposed in refs.~\cite{BW63,TY63}. For a recent overview of recent applications see e.g. refs.~\cite{Wei07,FG07} and references therein.} :
\begin{equation}\lbl{MBF}
\frac{x^2 (1-x)}{x^2 +\frac{\mathsf{t}}{m_{\mu}^2}(1-x)}=\frac{1}{2\pi i}\int\limits_{c_s-i\infty}^{c_s+i\infty}ds\left(\frac{4m_e^2}{\mathsf{t}} \right)^{s}
\left(\frac{4m_e^2}{m_{\mu}^2} \right)^{-s}x^{2s}(1-x)^{1-s}\ \Gamma(s)\Gamma(1-s)\,,
\end{equation}
where the threshold scale $4m_e^2$ is chosen for later convenience, 
and~\footnote{The Mellin variable $t$  should not be confused with the dispersive invariant mass squared $\mathsf{t}$.} 
\begin{equation}\lbl{MBL}
\log\left[1+\frac{x^2}{1-x}z(1-z)\frac{m_{\mu}^2}{m_{\tau}^2} \right]=	\frac{1}{2\pi i}\int\limits_{c_t-i\infty}^{c_t+i\infty}dt\left(\frac{m_\mu^2}{m_\tau^2} \right)^{-t}\left[\frac{x^2}{1-x}z(1-z)\right]^{-t}\frac{\Gamma(t)}{t}\Gamma(1-t)\,,
\end{equation}
where the integration paths along the imaginary axis are defined in the {\it fundamental strips}~\cite{FGD95}: 
\be
c_s =\Ree (s) \in\ \rbrack 0,1\lbrack \qquad \annd\qquad c_t =\Ree (t) \in \ \rbrack -1,0\lbrack\,.
\ee 
The interest of these representations lies in the property that the dependence on the physical mass ratios $\frac{4m_e^2}{m_{\mu}^2}$ and $\frac{m_\mu^2}{m_\tau^2}$ is then fully factorized and the Feynman parametric integrals one is left with are then those of a massless case. The only new feature is that they have to be computed as functions of the Mellin $s$--complex variable, in the case of electron and muon loops only; and as functions of the $(s\,,t)$--complex manifold in the case of electron, muon and tau loops. In all the cases we shall be concerned with here, it is possible to obtain analytic expressions for these functions in a rather straightforward way. In fact, they turn out to be rational functions of products of Gamma functions and Polygamma functions which depend linearly on the Mellin variables $s$, or $s$ and $t$.

Let us first discuss the case where we only have one  Mellin $s$--complex variable. It corresponds to the Feynman diagrams where only two mass scales are involved (i.e. the case where $l=\mu$ in Eq.~\rf{rep}). Using the representation in Eq.~\rf{MBF} one can rewrite Eq.~\rf{rep}, for a fixed $p$ and $j$, as follows
\be\lbl{amuimt}
a_{\mu}=\frac{\alpha}{\pi}\frac{1}{2\pi i}\int\limits_{c_s-i\infty}^{c_s+i\infty}ds
\left(\frac{4m_e^2}{m_{\mu}^2} \right)^{-s}\cM(s)\,,
\ee
with
\be\lbl{mellins}
\cM(s)=
\Gamma(s)\Gamma(1-s)  
\int_0^1 dx\ x^{2s}(1-x)^{1-s}\ F_{(j,p)}\  	\left[\Pi^{(\mu)}\left(\frac{-x^2}{1-x}m_\mu^2\right)\right]^p 
\int_{0}^{\infty}\frac{d\mathsf{t}}{\mathsf{t}}\left(\frac{4m_e^2}{\mathsf{t}} \right)^{s}\rho_{j}\left(\frac{4m_e^2}{\mathsf{t}}\right)\,.
\ee
The {\it converse mapping theorem} relates the asymptotic behaviour of  $a_{\mu}$ as a function of the small mass ratio ${4m_e^2}/{m_{\mu}^2}$, to the singularities of the integrand $\cM(s)$ as a function of the Mellin $s$--complex variable. For $m_e^2 \ll m_{\mu}^2$ the appropriate $s$--singularities are those in the left--hand--side of the {\it fundamental strip} and they are all on the negative real axis. The precise relation goes as follows: with   $\mathsf{p}\in\mathbb{R}$ and $k\in\mathbb{N}$, the function $\cM(s)$ in the l.h.s. of the {\it fundamental strip} has  a singular expansion of the type (ordered in increasing values of $\mathsf{p}$):
\be
\cM(s)\asymp\sum_{\mathsf{p}}\sum_{k}\frac{	\mathsf{a}_{\mathsf{p},k}}{(s+\mathsf{p})^{k+1}}\,.
\ee
The corresponding asymptotic behaviour of $a_{\mu}$ (ordered in increasing powers of $\mathsf{p}$) is then:
\begin{equation}
	a_{\mu}\underset{{\frac{4m_{e}^2}{m_{\mu}^2}\ \ra\  0}}{\thicksim}\frac{\alpha}{\pi}\sum_{\mathsf{p}}\sum_{k} \frac{(-1)^{k}}{k!}\ \mathsf{a}_{\mathsf{p},k}\ \left(\frac{4m_{e}^2}{m_{\mu}^2}\right)^{\mathsf{p}}\ \log^{k}\frac{4m_{e}^2}{m_{\mu}^2}\,,
\end{equation}
and the problem is then reduced to the calculation of the  $\mathsf{a}_{\mathsf{p},k}$ residues. This, in turn, requires the evaluation of the moment integrals
\be\lbl{moments}
\int_{0}^{\infty}\frac{d\mathsf{t}}{\mathsf{t}}\left(\frac{4m_e^2}{\mathsf{t}} \right)^{s}\rho_{j}\left(\frac{4m_e^2}{\mathsf{t}}\right)\quad\annd\quad 
\int_0^1 dx\ x^{2s}(1-x)^{1-s}\ 	\left[\Pi^{(\mu)}\left(\frac{-x^2}{1-x}m_\mu^2\right)\right]^p\,,
\ee
which appear in Eq.~\rf{mellins}

Let us consider a little bit further these two types of moments. First, the moments of the spectral functions $\rho_{j}\left(\frac{4m_e^2}{\mathsf{t}}\right)$ defined in Eqs.~\rf{dr1} to \rf{dr4} i.e.,
\begin{equation}
	\left(\frac{\alpha}{\pi}\right)^j R_{j}(s) \equiv \int_{0}^{\infty}\frac{d\mathsf{t}}{\mathsf{t}}\left(\frac{4m_e^2}{\mathsf{t}} \right)^{s}\rho_{j}\left(\frac{4m_e^2}{\mathsf{t}}\right)\,.
\end{equation}
One can immediately see that these moments can also be seen as  the Mellin transform of the spectral functions $\rho_{j}$ with respect to the variable $\xi=\frac{4m_e^2}{\mathsf{t}}$, since
\begin{equation}\lbl{spectralmellin}
	\left(\frac{\alpha}{\pi}\right)^j R_{j}(s)= \int_{0}^{\infty}d\xi\ \xi^{s-1}\ \rho_{j}(\xi)\,,
\qquad\xi=\frac{4m_e^2}{\mathsf{t}}\,,
\end{equation}
where in fact the integral over $\xi$ only runs from zero to one, because of the threshold factor  $\theta(\mathsf{t}-4m_e^2)$ in the lowest order spectral function in Eq.~\rf{losf}.
The interest of this point of view lies in the fact that the Mellin--Barnes representation relates the singular expansion of any of these $R_{j}(s)$ moments to  the asymptotic expansion of the spectral functions $\rho_{j}(\xi)$. Indeed, from Eq.~\rf{spectralmellin} and the asymptotic behaviour of $\rho_{j}(\xi)$ at small $\xi$ and at large $\xi$, there follows that
\begin{equation}
	\rho_{j}(\xi)=	\left(\frac{\alpha}{\pi}\right)^j\frac{1}{2\pi i}\int\limits_{r_s-i\infty}^{r_s+i\infty}ds\ \left(\frac{1}{\xi} \right)^{s} 
	 \ R_{j}(s)\,,\quad {\rm with}\quad r_s =\Ree (s) \in\ \rbrack 0,\infty\lbrack\,.
\end{equation}
In other words, the {\it direct mapping} (in the sense of ref.~\cite{FGD95}) establishes a precise relation between  the asymptotic expansion of the spectral functions $\rho_{j}(\xi)$ at small $\xi$ (which  is known from their explicit analytic expression) and the singular series of the moments  $R_{j}(s)$, for $s\le 0$, that we are interested in. We shall come back later on to these relations in somewhat more detail.
 
 The second type of moments in Eq.~\rf{moments} are also well defined Mellin transforms; in this case with respect to the invariant photon momenta
 \be
\omega\equiv\frac{Q^2}{m_{\mu}^2}=\frac{x^2}{1-x}\,,
\ee
flowing in the basic muon vertex diagram. Indeed, 
one can easily verify that

\bea\lbl{xandrho}
	\left(\frac{\alpha}{\pi}\right)^p \Omega_{p}(s)\equiv \lefteqn{ \int_0^1 dx\ x^{2s}(1-x)^{1-s}\ 	\left[\Pi^{(\mu)}\left(\frac{-x^2}{1-x}m_\mu^2\right)\right]^p=} \nn \\ & & \int_0^\infty d\omega\ 
 \omega^{s-1} \sqrt{\frac{\omega}{4+\omega}}\left(\frac{\sqrt{4+\omega}-\sqrt{\omega}}{ \sqrt{4+\omega}+\sqrt{\omega}} \right)^2\ \left[\Pi^{(\mu)}(-\omega\ m_\mu^2)\right]^p \,.
\eea

\noi
Again, the evaluation of the singular series associated to the moments $\Omega_{p}(s)$, for $s\le 0$,  is very much facilitated by the {\it direct mapping} which relates them to the asymptotic expansion of $\left[\Pi^{(\mu)}(-\omega\ m_\mu^2)\right]^p $ for small $\omega$. 

The remarkable property of the Mellin--Barnes representation is precisely the factorization in terms of moment integrals as shown in Eq.~\rf{mellins}. It is in fact this factorization which is at the basis of the renormalization group properties  discussed  by many authors. The algebraic factorization above, however, is more general and it also shows the full underlying renormalization group structure at work. The classical renormalization group constrains which have been exploited in the literature only apply to the evaluation of {\it leading} asymptotic behaviours ({\it powers of logarithms and constant terms}). In the example above, this is encoded by the properties of the Mellin singularity at $s=0$. This singularity governs the spectral function moments
$R_j(0)$ on the one hand, as well as the contribution  to the muon anomaly from vacuum polarization muon loop insertions ($p$ loops): the moments $\Omega_p (0)$. Notice that the moments $R_j(0)$ are      
UV--singular, hence the need of the $s$-- regularization.  The singularity is related to charge renormalization and therefore to the QED $\beta$--function at a given order in perturbation theory~\cite{LdeR74}, in our case the lowest order. The predictive power of the renormalization group structure lies in the fact that once we know the two types of moments in Eq.~\rf{moments} at a given order in perturbation theory we have a prediction at a higher order for the convolution which in our case gives $a_{\mu}$ in Eq.~\rf{amuimt}. What is new here is that this factorization extends as well to the subleading terms in the expansion modulated by inverse ${{m_{\mu}^2}/m_e^2}$--powers. What governs these terms is nothing but the residues of the successive Mellin singularities in the negative real axis. Here, the expansion in inverse ${{m_{\mu}^2}/m_e^2}$--powers is analogous to the $1/Q^2$--expansion in the Operator Product Expansion of Green's functions in QCD, while the moments $R_j(s)$, for $s<0$, are the equivalent of the so called vacuum condensates. The big difference, of course, is that in QED these condensates are moments of spectral functions which are known at all values of $\mathsf{t}$ and, therefore, can be calculated explicitly at a given order in perturbation theory. Notice, however, that also in QED these moments (condensates) are {\it a priori} ill defined because they are singular. The Mellin--Barnes technique provides a systematic regularization and hence a precise separation of short--distance  and long--distance effects.   

The case of two Mellin $(s\,,t)$--complex variables, corresponding to Feynman diagrams with both  electron loops and  tau loops, is conceptually more complicated and it requires a specific treatment which we shall provide in Section~VII. A detailed discussion of the very interesting underlying mathematics which governs this case will be given in a forthcoming publication~\cite{AGdeR08}.

We finally give below the exact explicit expressions, in the Mellin--Barnes integral representation,  of the contribution to the muon anomaly from each of the specific set of Feynman diagrams in Fig.~1 and Fig.~2. 

%%%%%%%%%%%%%%%%%%%%%%%%%%%%%%%%%%%%%%%%%%%%%%%%%%%%%%%%%%%%%%%%%%%%%%%%
%%%%%%%%%%%%%%%%%%%%%%%%%%%%%%%%%%%%%%%%%%%%%%%%%%%%%%%%%%%%%%%%%%%%%%%%
\subsection{\small Eighth Order Contributions}\lbl{eigth}
%\setcounter{equation}{0}
%\def\theequation{\arabic{section}.\arabic{equation}} 
%%%%%%%%%%%%%%%%%%%%%%%%%%%%%%%%%%%%%%
\begin{figure}[h]

\begin{center}
\includegraphics[width=0.6\textwidth]{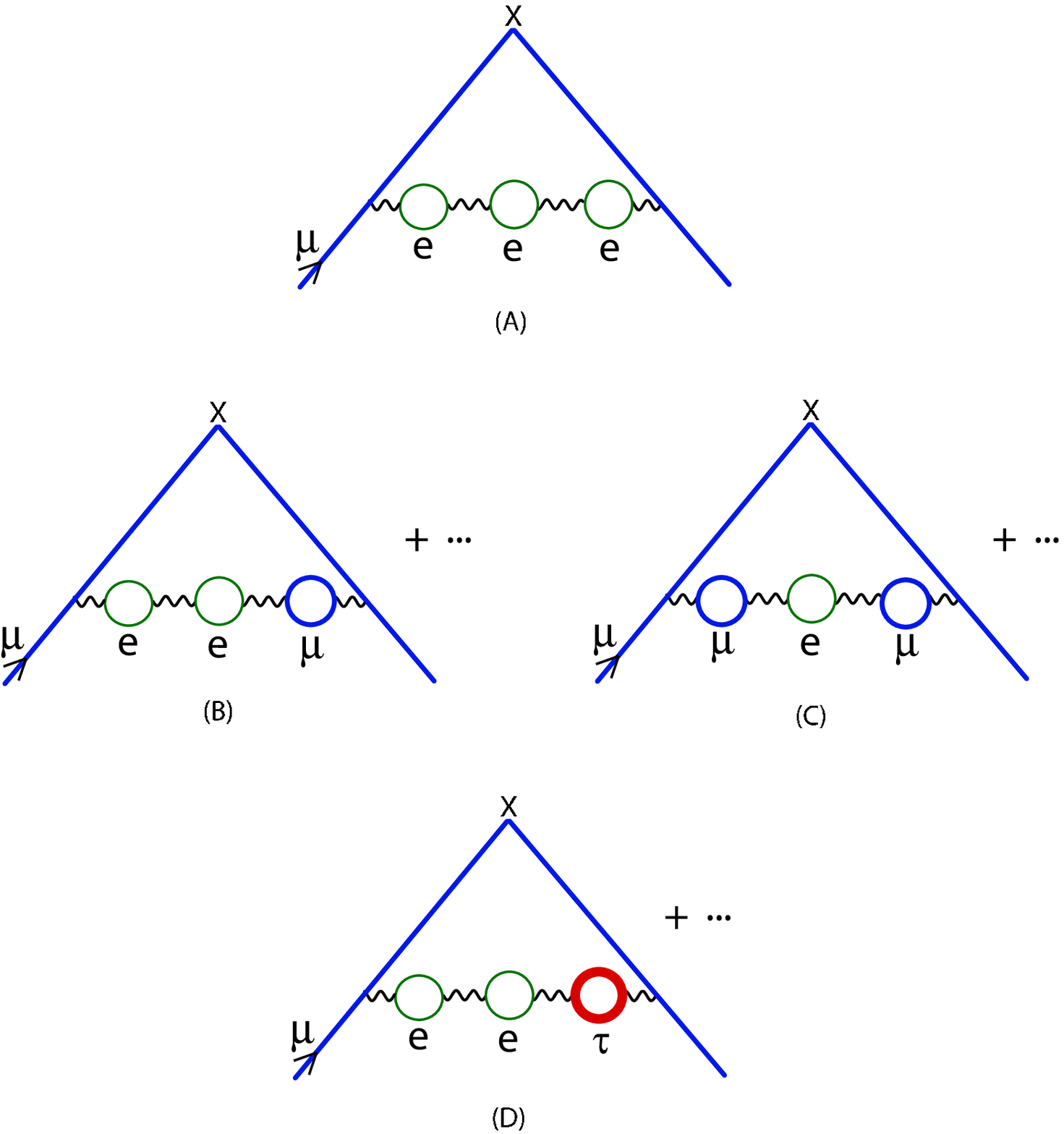}

\end{center}

\vspace*{0.25cm}
{\bf Fig.~1}
{\it\small  Eighth--order Feynman diagrams with lowest order vacuum polarization electron--loops and a $\tau$--loop which contribute to the Muon Anomaly and are  enhanced by powers of $\log\frac{m_{\mu}}{m_e}$ factors. The dots indicate the other diagrams with different permutations of the lepton--loops.
}

\end{figure}
%%%%%%%%%%%%%%%%%%%%%%%%%%%%%%%%%%%%%% 

\begin{itemize}
	\item {\sc Three Electron Loops}, Fig.~1(A) [one diagram]:

\be\lbl{eee}
	a_{\mu}^{(eee)}  =  \left(\frac{\alpha}{\pi}\right)^4 \frac{\bf 1}{2\pi i}\int\limits_{c_s-i\infty}^{c_s+i\infty}ds \left(\frac{4m_e^2}{m_{\mu}^2} \right)^{-s}\Gamma(s)\Gamma(1-s)\ \Omega_0 (s)\ R_{3}(s)\,.
\ee

	\item {\sc Two Electron Loops and One Muon Loop}, Fig.~1(B) [three diagrams]:

\be\lbl{eemu}
	a_{\mu}^{(ee\mu)}  =  \left(\frac{\alpha}{\pi}\right)^4 \ \frac{ {\bf 3}}{2\pi i}\int\limits_{c_s-i\infty}^{c_s+i\infty}ds \left(\frac{4m_e^2}{m_{\mu}^2} \right)^{-s}\Gamma(s)\Gamma(1-s)\ \Omega_1 (s)\ R_2 (s)\,.
\ee

	\item {\sc One Electron Loop and Two Muon Loops}, Fig.~1(C) [three diagrams]:

\be\lbl{emumu}
	a_{\mu}^{(e\mu\mu)}  =  \left(\frac{\alpha}{\pi}\right)^4 \ \frac{{\bf 3}}{2\pi i}\int\limits_{c_s-i\infty}^{c_s+i\infty}ds \left(\frac{4m_e^2}{m_{\mu}^2} \right)^{-s}\Gamma(s)\Gamma(1-s)\ \Omega_2 (s)\ R_1 (s) \,.
\ee

\noi

	\item {\sc Two Electron Loops and One Tau Loop}, Fig.~1(D) [three diagrams]:
	
	Using the representation in Eq.~\rf{rtau}for the tau self--energy function as well as  the Mellin--Barnes representation in Eq.~\rf{MBL} one easily gets the expression

{\setl
\bea\lbl{eetau}
\hspace*{-2cm}	a_{\mu}^{(ee\tau)} & = & \left(\frac{\alpha}{\pi}\right)^4 \frac{{\bf 3}}{2\pi i} \int\limits_{c_s-i\infty}^{c_s+i\infty} ds \left(\frac{4m_e^2}{m_{\mu}^2} \right)^{-s}\frac{1}{2\pi i}\int\limits_{c_t-i\infty}^{c_t+i\infty} dt\left(\frac{m_\mu^2}{m_\tau^2} \right)^{-t}\times \nn \\
& & 	 \Gamma(s)\Gamma(1-s)\ \frac{\Gamma(t)}{t}\Gamma(1-t)\ \Theta(s,t)\  R_2 (s)\,, 
\eea}

\noi
where $\Theta(s,t)$ is the Feynman parametric integral (see Eqs.~\rf{MBF} and \rf{MBL})

{\setl
\bea\lbl{theta}
\Theta(s,t) & =  &\int_0^1 dx x^{2s}(1-x)^{1-s} \int_0^1 dz 
\left(\frac{x^2}{1-x}z(1-z)\right)^{-t} \nn \\
 & = & (-2)\ \frac{\Gamma(2-t)\Gamma(2-t)}{\Gamma(4-2t)}\ \frac{\Gamma(1+2s-2t)\Gamma(2-s+t)}{\Gamma(3+s-t)}\,.
\eea}

\noi
 Notice that the dependence on the variables $s$ and $t$ in the function $\Theta(s,t) $ is not   factorized. It is this fact that requires new technical considerations which we shall discuss in Section~VII.

\end{itemize}

%%%%%%%%%%%%%%%%%%%%%%%%%%%%%%%%%%%%%%%%%%%%%%%%%%%%%%%%%%%%%%%%%%%%%%%%
%%%%%%%%%%%%%%%%%%%%%%%%%%%%%%%%%%%%%%%%%%%%%%%%%%%%%%%%%%%%%%%%%%%%%%%%
\subsection{\small Tenth Order Contributions}\lbl{tenth}
%\setcounter{equation}{0}
%\def\theequation{\arabic{section}.\arabic{equation}} 
%%%%%%%%%%%%%%%%%%%%%%%%%%%%%%%%%%%%%%
%%%%%%%%%%%%%%%%%%%%%%%%%%%%%%%%%%%%%%
\begin{figure}[h]

\begin{center}
\includegraphics[width=0.6\textwidth]{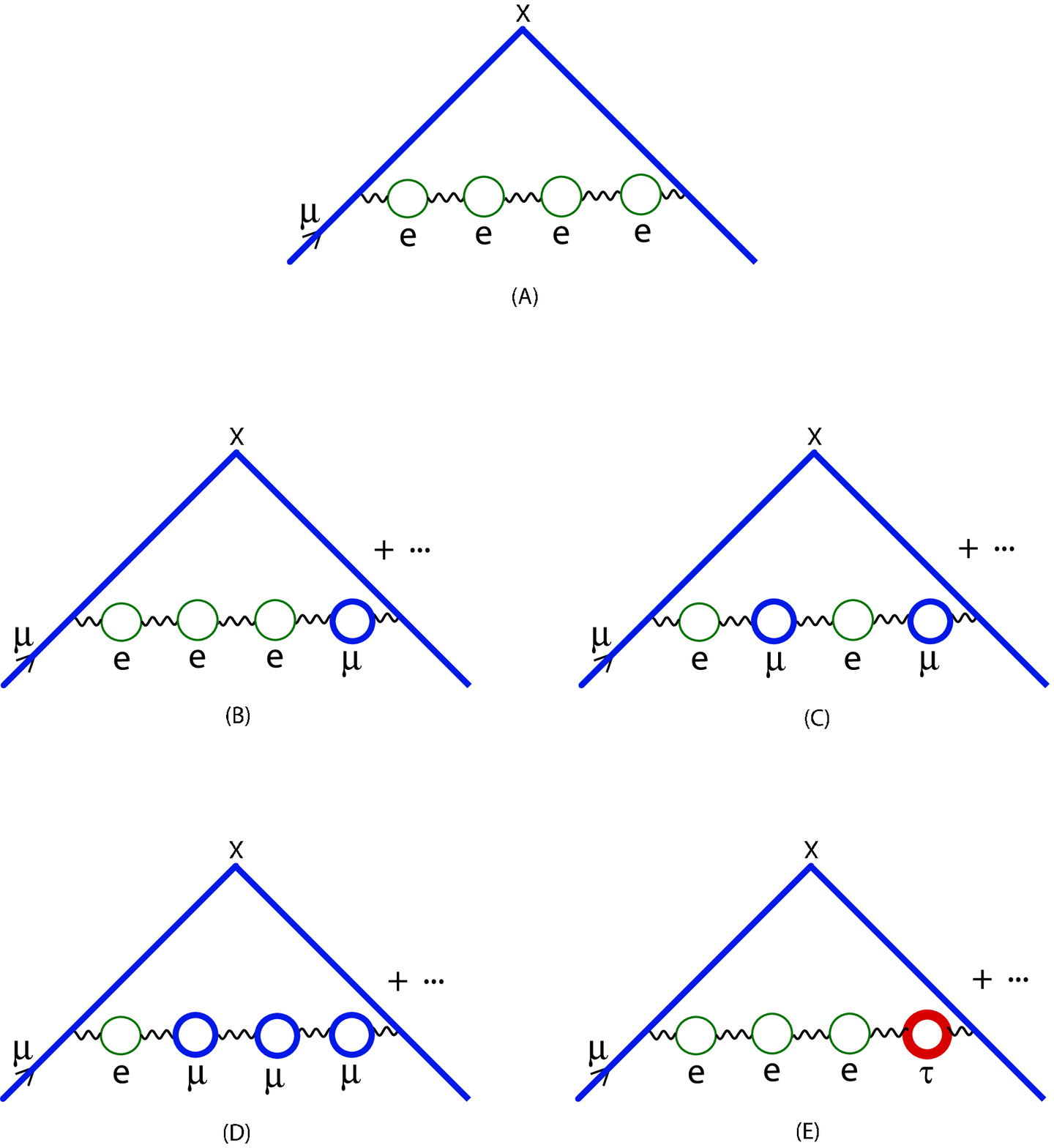}

\end{center}

\vspace*{0.25cm}
{\bf Fig.~2}
{\it\small  Tenth--order Feynman diagrams with lowest order vacuum polarization electron--loops and a $\tau$--loop which contribute to the Muon Anomaly and are  enhanced by powers of $\log\frac{m_{\mu}}{m_e}$ factors. The dots indicate the other diagrams with different permutations of the lepton--loops.
}

\end{figure}
%%%%%%%%%%%%%%%%%%%%%%%%%%%%%%%%%%%%%% 

\begin{itemize}
	\item {\sc Four Electron Loops}, Fig.~2(A) [one diagram]:

\be\lbl{eeee}
	a_{\mu}^{(eeee)}  =  \left(\frac{\alpha}{\pi}\right)^5\  \frac{(-{\bf 1})}{2\pi i}\int\limits_{c_s-i\infty}^{c_s+i\infty}ds \left(\frac{4m_e^2}{m_{\mu}^2} \right)^{-s}\Gamma(s)\Gamma(1-s)\Omega_{0}(s)\ R_4 (s)\,.
\ee

	\item {\sc Three Electron Loops and One Muon Loop}, Fig.~2(B) [four diagrams]:
	
\be\lbl{eeemu}
	a_{\mu}^{(eee\mu)}  =  \left(\frac{\alpha}{\pi}\right)^5  (-{\bf 4})\ \frac{(-{\bf 4})}{2\pi i}\int\limits_{c_s-i\infty}^{c_s+i\infty}ds \left(\frac{4m_e^2}{m_{\mu}^2} \right)^{-s}\Gamma(s)\Gamma(1-s)\ \Omega_1 (s)\ R_3 (s)\,.
\ee

	\item {\sc Two Electron Loops and Two Muon Loops}, Fig.~2(C) [six diagrams]:
	
	\be\lbl{eemumu}
	a_{\mu}^{(ee\mu\mu)}  =  \left(\frac{\alpha}{\pi}\right)^5 \ \frac{(-{\bf 6})}{2\pi i}\int\limits_{c_s-i\infty}^{c_s+i\infty}ds \left(\frac{4m_e^2}{m_{\mu}^2} \right)^{-s}\Gamma(s)\Gamma(1-s)\ \Omega_2 (s)\ R_2 (s)\,.
\ee
\noi

	\item {\sc One Electron Loop and Three Muon Loops}, Fig 2.(D) [four diagrams]:
	
\be\lbl{emumumu}
	a_{\mu}^{(e\mu\mu\mu)}  = \left(\frac{\alpha}{\pi}\right)^5 \ \frac{(-{\bf 4})}{2\pi i}\int\limits_{c_s-i\infty}^{c_s+i\infty}ds \left(\frac{4m_e^2}{m_{\mu}^2} \right)^{-s}\Gamma(s)\Gamma(1-s)\ \Omega_3 (s)\ R_1 (s) \,.
\ee

	\item {\sc Three Electron Loops and One Tau Loop}, Fig.~2(E) [four diagrams]:
	
	The only change here, with respect to the representation in Eq.~\rf{eetau}, is the convolution with the moment $R_{3}(s)$ (instead of $R_{2}(s)$) and the  combinatorial factor $(-{\bf 4})$ [instead of $({\bf 3})$]:

{\setl
\bea\lbl{eeetau}
\hspace*{-2cm}	a_{\mu}^{(eee\tau)} & = & \left(\frac{\alpha}{\pi}\right)^5 \frac{(-{\bf 4})}{2\pi i} \int\limits_{c_s-i\infty}^{c_s+i\infty} ds \left(\frac{4m_e^2}{m_{\mu}^2} \right)^{-s}\frac{1}{2\pi i}\int\limits_{c_t-i\infty}^{c_t+i\infty} dt\left(\frac{m_\mu^2}{m_\tau^2} \right)^{-t}\times \nn \\
	& &  \Gamma(s)\Gamma(1-s)\ \frac{\Gamma(t)}{t}\Gamma(1-t)\ \Theta (s,t)\  R_3 (s)\,, 
\eea}

\noi
where $\Theta (s,t)$ is the function defined in Eq.~\rf{theta}.
\end{itemize}

%%%%%%%%%%%%%%%%%%%%%%%%%%%%%%%%%%%%%%%%%%%%%%%%%%%%%%%%%%%%%%%%%%%%%%%%
%%%%%%%%%%%%%%%%%%%%%%%%%%%%%%%%%%%%%%%%%%%%%%%%%%%%%%%%%%%%%%%%%%%%%%%%
%%%%%%%%%%%%%%%%%%%%%%%%%%%%%%%%%%%%%%%%%%%%%%%%%%%%%%%%%%%%%%%%%%%%%%%%
%%%%%%%%%%%%%%%%%%%%%%%%%%%%%%%%%%%%%%%%%%%%%%%%%%%%%%%%%%%%%%%%%%%%%%%%
\section{\small The Moment Integrals}\lbl{paramints}
\setcounter{equation}{0}
\def\theequation{\arabic{section}.\arabic{equation}}

We are now left with two types of moment integrals: the $R_j (s)$ moments of the electron--positron spectral functions in Eq.~\rf{spectralmellin}, (the integrals over the variable $\xi$, or equivalently the variable $\delta$ in Eq.~\rf{delta}); and the $\Omega_p (s)$ moments of the euclidean muon vacuum polarization, (the integrals over the invariant photon momenta $\omega$, or equivalently the Feynman $x$--parameter). It is a luxury that these integrals can all be done analytically. Here we shall give the results and comment on a few technical points; in particular on their relation to the underlying renormalization group properties.

\begin{itemize}

	\item {\sc The $\Omega_0 (s)$ Moment.}
	
	This  moment, which appears in $a_{\mu}^{(eee)}$ and $a_{\mu}^{(eeee)}$,  corresponds to the  trivial $x$--integral
\begin{equation}\lbl{omega0}
	\Omega_0 (s)=\int_0^1 dx (1-x) \left(\frac{x^2}{1-x}\right)^{s}=\frac{\Gamma(1+2s)\Gamma(2-s)}{\Gamma(3+s)}\,.
\end{equation}
The particular value $\Omega_0 (0)=\frac{1}{2}$ is precisely the coefficient of the lowest order muon anomaly, the Schwinger term: $a_{\mu}=\left(\frac{\alpha}{\pi}\right)\left[\Omega_0 (0)=\frac{1}{2}\right]$. Notice that $\Omega_0 (s)$ is  IR--singular at $s=-1/2, -3/2, \cdots$. It is precisely these IR--singularities that are at the origin	of the contributions in odd powers of $m_e /m_{\mu}$ in $a_{\mu}^{(e)}$, $a_{\mu}^{(ee)}$, $a_{\mu}^{(eee)}$ and $a_{\mu}^{(eeee)}$ that one encounters in the explicit calculations.

	\item {\sc The $R_1 (s)$ Moment.}

This moment appears in the expressions for $a_{\mu}^{(e\mu\mu)}$ and $a_{\mu}^{(e\mu\mu\mu)}$ in Eqs.~\rf{emumu} and \rf{emumumu} and it corresponds to the integral
	\be
	R_1 (s)=\int_0^1 d\delta\ 2\delta (1-\delta^2)^{s-1}\ \frac{1}{2}\delta \left(1-\frac{1}{3}\delta^2 \right)\,,
	\ee
which can be trivially computed, with the simple result:
\begin{equation}\lbl{sigma1}
	R_1 (s)=\frac{\sqrt{\pi}}{4}\ \frac{1}{s}  \ \frac{\Gamma(2+s)}{\Gamma\left(\frac{5}{2}+s\right)}\,,	
\end{equation}
explicitly showing the singularities in the negative real axis (the origin $s=0$ included).
Notice that, because of the zeros provided by the factor $\frac{1}{\Gamma\left(\frac{5}{2}+s\right)}$, there cannot be terms of $\cO\left[\left(\frac{m_e^2}{m_{\mu}^2}\right)^p \right]$ for $p= 5/2+n$, $n=0,1,2,\cdots$ in vacuum polarization contributions to $a_{\mu}$ from one electron loop. 

\end{itemize}

\noi
The two moments $\Omega_0 (s)$ and $R_1 (s)$  are the only quantities required to evaluate  the contribution to $a_{\mu}^{(e)}$ from the fourth order Feynman diagram in Fig.~3, which, in our representation, is given by the integral
\begin{equation}\lbl{e}
	a_{\mu}^{(e)}  =  \left(\frac{\alpha}{\pi}\right)^2 \frac{1}{2\pi i}\int\limits_{c_s-i\infty}^{c_s+i\infty}ds \left(\frac{4m_e^2}{m_{\mu}^2} \right)^{-s}\Gamma(s)\Gamma(1-s)\ \Omega_0(s)\ R_{1}(s)\,.
\end{equation}

%%%%%%%%%%%%%%%%%%%%%%%%%%%%%%%%%%%%%%
\begin{figure}[h]

\begin{center}
\includegraphics[width=0.3\textwidth]{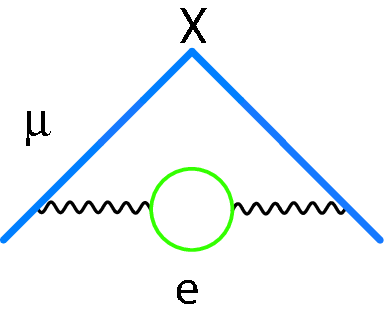}

\end{center}

\vspace*{0.25cm}
{\bf Fig.~3}
{\it\small  Fourth--order Feynman diagram with a lowest order vacuum polarization electron--loop.
}

\end{figure}
%%%%%%%%%%%%%%%%%%%%%%%%%%%%%%%%%%%%%%

\noi
Since the moments  $\Omega_0(s)$ and $R_{1}(s)$ are  known analytically, the singular series expansion of the integrand can be computed up to as many terms as one wishes; e.g.,

{\setl
\bea\lbl{emellin}
\Gamma(s)\Gamma(1-s)\ \Omega_0(s)\ R_{1}(s) & \asymp &  \frac{1}{6}\frac{1}{s^2}+\left(\frac{1}{3}\log 2 -\frac{25}{36} \right)\frac{1}{s} \nn \\
 & & +\frac{\pi^2}{8}\frac{1}{1/2+s} \nn \\
 & & -\frac{1}{2}\frac{1}{(1+s)^2}+\left(\frac{3}{4}-\log 2\right)\frac{1}{1+s} +\cdots \,,
\eea}

\noi
and the {\it converse mapping theorem} tells us how to read in a straightforward way the corresponding asymptotic expansion contribution~\footnote{The full asymptotic expansion can be found in ref.~\cite{FGdeR05} where references to earlier calculations are also given.}:

{\setl
\bea\lbl{easymp}
	a_{\mu}^{(e)}\underset{{\frac{m_{e}^2}{m_{\mu}^2}\ \ra\  0}}{\thicksim} &  &  \left(\frac{\alpha}{\pi}\right)^2 \left\{\frac{1}{6}\log\frac{m_\mu^2}{m_e^2}  -\frac{25}{36} \right. \nn \\
 & & +\frac{\pi^2}{4}\left(\frac{m_e^2}{m_\mu^2}\right)^{1/2} \nn \\
 & & 
 \left.  + \frac{m_e^2}{m_{\mu}^2} \left[-2\log\frac{m_\mu^2}{m_e^2}  +3 \right]  + \cO\left[\left(\frac{m_e^2}{m_\mu^2}\right)^{3/2} \right] \right\}\,.
\eea}

\noi
The reason why we consider this well known contribution here is that it allows us to discuss the factorization properties of the Mellin--Barnes representation and its relation to the underlying renormalization group structure in a case which, in spite of its simplicity, illustrates the generic features rather well. Notice that the singular expansion in Eq.~\rf{emellin} results from the combination of the Laurent series of three factors: the geometric series factor $\Gamma(s)\Gamma(1-s)$, which appears in the original Mellin--Barnes representation in Eq.~\rf{MBF}, and the two moments $\Omega_0(s)$ and  $R_{1}(s)$. The residue of the leading singularity in $1/s^2$, which provides the coefficient of the $\log\frac{m_\mu^2}{m_e^2}$ term in Eq.~\rf{easymp},  can be read off directly from the asymptotic behaviour of the lowest order spectral function $\rho\left(\frac{4m_e^2}{\mathsf{t}}\right)\underset{{\mathsf{t}\ra\infty}}{\thicksim}\frac{1}{3}$, (which in turn is correlated to the residue of the leading term in the singular expansion of $R_{1}(s)\asymp \frac{1}{3}\frac{1}{s}$) and the lowest order result $\Omega_{0}=1/2$~; the geometric factor $\Gamma(s)\Gamma(1-s)$ providing the extra $1/s$ factor. This is precisely the {\it leading} prediction of the  renormalization group in this case ~\cite{LdeR74}. 

The next--to--next--to--leading term in the asymptotic expansion in Eq.~\rf{easymp} (the second line) is governed by the $\frac{1}{s+1/2}$ term in the singular expansion in Eq.~\rf{emellin} (the second line in this equation). The origin of this singularity is the moment $\Omega_0(s)$, which is singular at $s=-1/2$; the other factors are regular: $\Gamma(-1/2)\Gamma(1+1/2)=-\pi$ and  $R_{1}(1/2)=-\frac{\pi}{4}$. Again, the residue of the $\frac{1}{s+1/2}$--singularity of $\Omega_0(s)$ can be read off from the leading term in the asymptotic expansion of the integrand function in Eq.~\rf{xandrho} i.e.,
\be
\sqrt{\frac{\omega}{4+\omega}}\left(\frac{\sqrt{4+\omega}-\sqrt{\omega}}{ \sqrt{4+\omega}+\sqrt{\omega}} \right)^2 \underset{{\omega\ra\ 0}}{\thicksim} \frac{1}{2}\ \sqrt{\omega}+\cO(\omega)\,.
\ee
We see, therefore,  how the factorization of the Mellin--Barnes representation  allows one to fix a non--trivial coefficient of the asymptotic expansion of an $\cO\left[\left(\frac{\alpha}{\pi}\right)^2\right]$ quantity from the knowledge of two numbers which appear at a lower  $\cO\left(\frac{\alpha}{\pi}\right)$ and have been very easily identified.

What about the next--to--leading term? This refers to the factor $-\frac{25}{36}$ in the first line of Eq.~\rf{easymp}, which is not fixed by simple renormalization group arguments. The reason why  simple renormalization group arguments fail to fix this term is due to the fact that it does not originate from products of leading terms of singular expansions only. More precisely, the coefficient of the $\frac{1}{s}$ term in Eq.~\rf{emellin} also depends on the next--to--leading terms of the Laurent series of each of the individual factors $\Omega_0(s)\,,$ and  $R_{1}(s)$. These terms, however, can also be easily obtained without having to calculate  explicitly the functions $\Omega_0(s)$ and  $R_{1}(s)$. Indeed, since $\Omega_0(s)$ is regular at $s\ra 0$ we can simply Taylor expand the integrand and find
\be\lbl{taylorO}
\Omega_0(s) \underset{{s\ra\ 0}}{\thicksim} \frac{1}{2}-\frac{5}{4}s
+\cO(s^2)\,.
\ee
By contrast,  the moment $R_{1}(s)$ is singular at $s\ra 0$; but here we know that the residue of the singularity originates in the leading term of the asymptotic expansion of the lowest order spectral function in the integrand. Therefore, subtracting this leading term from the spectral function itself produces a regular integral from which one can extract the regular series in the Laurent expansion at $s\ra 0$ by simple Taylor expansion. The only integral we have to do  is then
\be\lbl{TaylorS1}
 	\left(\frac{\alpha}{\pi}\right) R_{1}(s)\underset{{s\ra\ 0}}{\thicksim} \frac{1}{3s} +\int_{0}^{\infty}d\xi\ \xi^{-1}\ \left(\rho_{1}(\xi)-\frac{1}{3}\right)+ \cO(s)=\frac{1}{3s} +\frac{1}{9}(6\log 2 -5)+ \cO(s) \,.
 	\ee
Combining the results in Eqs.~\rf{taylorO} and \rf{TaylorS1} with the fact that
$
\Gamma(s)\Gamma(1-s)\underset{{s\ra\ 0}}{\thicksim} \frac{1}{s}+\cO(s)\,, 
$
one easily gets  the coefficient of the $\frac{1}{s}$--term in Eq.~\rf{emellin} and hence the term $-\frac{25}{36}$ in the asymptotic expansion in Eq.~\rf{easymp}.

We leave to the studious reader the pleasure of reproducing the residues of the $\frac{1}{(s+1)^2}$ and  $\frac{1}{(s+1)}$ singularities in Eq.~\rf{emellin}, and hence the terms of $\cO\left(\frac{m_e^2}{m_\mu^2} \right)$ in Eq.~\rf{easymp}, using similar simple arguments.

\begin{itemize}
	\item {\sc The $R_2(s)$ Moment.}

This moment appears in the expressions of $a_{\mu}^{(ee\mu)}$, $a_{\mu}^{(ee\tau)}$ and $a_{\mu}^{(ee\mu\mu)}$ in Eqs.~\rf{eetau}, \rf{eemu} and \rf{eemumu} and it is given by the integral	 
\begin{equation}
		\left(\frac{\alpha}{\pi}\right)^2\ R_2 (s)=\int_{0}^{\infty} d\xi\  \xi^{s-1}\ \rho_2 (\xi)\,,
\end{equation}
associated to the spectral function in the r.h.s. of Eq.~\rf{dr2}. In terms of the variable  $\delta$, it gives rise to the integral
\begin{equation}
R_2 (s)=	2\ \int_0^1 d\delta\ 2\delta (1-\delta^2)^{s-1}\ \frac{1}{2}\delta \left(1-\frac{1}{3}\delta^2 \right)\left[\frac{8}{9}-\frac{1}{3}\delta^2+\delta\left(\frac{1}{2}-\frac{1}{6}\delta^2\right) 
\log\frac{1-\delta}{1+\delta}\right]\,,
\end{equation}
which can be done analytically rather easily with the result
\begin{equation}\lbl{sigma2}
R_2 (s)=\frac{\sqrt{\pi}}{9}\ \frac{(-1+s)( 6+13s+4s^2)}{s^2 (2+s)(3+s)}\ \frac{\Gamma(1+s)}{\Gamma\left(\frac{3}{2}+s\right)}\,,
\end{equation}
explicitly showing the singularities in the negative real axis.
Notice that, because of the zeros provided by the factor $\frac{1}{\Gamma\left(\frac{3}{2}+s\right)}$, there cannot be terms of $\cO\left[\left(\frac{m_e^2}{m_{\mu}^2}\right)^p \right]$ for $p= 3/2+n$, $n=0,1,2,\cdots$ in vacuum polarization contributions to $a_{\mu}$ from two electron loops.

	\item { \sc The $\Omega_1(s)$ Moment.}
	
It corresponds to the Mellin transform (see Eq.~\rf{xandrho})

\bea\lbl{xandrho1}
	\left(\frac{\alpha}{\pi}\right) \Omega_{1}(s)=\lefteqn{ \int_0^\infty d\omega\ 
 \omega^{s-1} \sqrt{\frac{\omega}{4+\omega}}\left(\frac{\sqrt{4+\omega}-\sqrt{\omega}}{ \sqrt{4+\omega}+\sqrt{\omega}} \right)^2\ \left[\Pi^{(\mu)}(-\omega\ m_\mu^2)\right]=}  \nn \\ & &
 \int_0^1 dx\ (1-x)\left(\frac{x^2}{1-x} \right)^s \ 	\left[\Pi^{(\mu)}\left(\frac{-x^2}{1-x}m_\mu^2\right)\right]
 \,.
\eea

This integral can also be easily done, using the expression in  Eq.~\rf{ree} in terms of a subtracted logarithm:
\be\lbl{sublog}
\log(1-x)\ra \left[\log(1-x)-\left(-x-\frac{x^2}{2}-\frac{x^3}{3}\right)\right]+\left(-x-\frac{x^2}{2}-\frac{x^3}{3}\right)\,. 
\ee
The integral over $x$ is then convergent at $x=0$ term by term, with the result

{\setl
\bea\lbl{omega1}
\Omega_{1}(s) & = & \Gamma(2-s)	\left\{ -\frac{4}{3}\frac{\Gamma(-1+2s)}{\Gamma(1+s)}+\frac{4}{3}\frac{\Gamma(2s)}{\Gamma(2+s)}+\frac{5}{9}\frac{\Gamma(1+2s)}{\Gamma(3+s)} \right.\nn \\
		 &  & +\left[-\frac{4}{3}\frac{\Gamma(-2+2s)}{\Gamma(s)}+2\frac{\Gamma(-1+2s)}{\Gamma(1+s)}
	 -\frac{1}{3}\frac{\Gamma(1+2s)}{\Gamma(3+s)}\right]\rH_{1-s} + \nn \\
	  & & \left.\frac{4}{3}\frac{\Gamma(-2+2s)}{\Gamma(s)}\rH_{-1+s}-2
\frac{\Gamma(-1+2s)}{\Gamma(1+s)}\rH_{s}+\frac{1}{3}\frac{\Gamma(1+2s)}{\Gamma(3+s)}\rH_{2+s}\right\}\,,
	  	 \eea}
	 
\noi
where $\rH_s$ denotes the function
\be
\rH_s = \psi(1+s)+\gamma_{\rm E}\,,
\ee
related to the $\psi$--function\footnote {Recall that the $\psi(z)$--function is meromorphic with simple poles at $z=0\,,-1\,, -2\dots $}~ $\psi(z)=\frac{d}{dz}\log\Gamma(z)$, as follows
\be
\psi(z) = -\gamma_E + \sum_{m=0}^{\infty} \left( \frac{1}{m+1} - \frac{1}{z+m}
\right)\,,
\ee
and, therefore, when $z$ is an integer $n$, $\rH_n$ corresponds to the Harmonic sum
\be
\rH_n =1+\frac{1}{2}+	\frac{1}{3}+\frac{1}{4}+\cdots +\frac{1}{n}\,.
\ee

Notice that $\Omega_{1}(0)=-\frac{119}{36}+\frac{\pi^2}{2}$
is precisely the well known coefficient of the fourth order contribution to $a_{\mu}^{(\mu)}$ from the Feynman graph in Fig.~4:
\be
a_{\mu}^{(\mu)}=\left(\frac{\alpha}{\pi} \right)^2 \left(-\frac{119}{36}+\frac{\pi^2}{2}\right)\,.
\ee
%%%%%%%%%%%%%%%%%%%%%%%%%%%%%%%%%%%%%%
\begin{figure}[h]

\begin{center}
\includegraphics[width=0.3\textwidth]{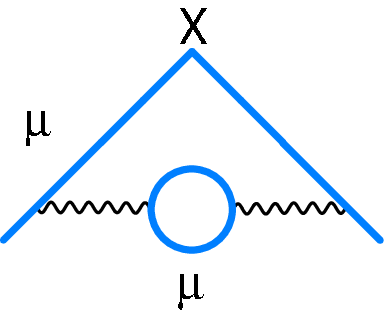}

\end{center}

\vspace*{0.25cm}
{\bf Fig.~4}
{\it\small  Fourth--order Feynman diagram with a lowest order vacuum polarization muon--loop.
}

\end{figure}
%%%%%%%%%%%%%%%%%%%%%%%%%%%%%%%%%%%%%%

Another property of $\Omega_{1}(s)$, which is also valid for all $\Omega_{p}(s)$ with $p\ge 1$, is the fact that they are no longer singular at $s=-1/2$. This can be readily seen from the fact that $\Pi^{(\mu)}\left(\frac{-x^2}{1-x}m_\mu^2\right)$ in Eq.~\rf{xandrho1} vanishes at $x=0$:
\be\lbl{selfmu0}
\Pi^{(\mu)}\left(\frac{-x^2}{1-x}m_\mu^2\right)\underset{{x\ra\ 0}}{\thicksim}\left(\frac{\alpha}{\pi}\right)\left[-\frac{1}{15}x^2 +\cO(x^3 )\right] \,.
\ee
This is why Feynman graphs with electron loops and at least one muon loop insertion in the vacuum polarization have no $\cO\left[\left(\frac{m_e^2}{m_\mu^2}\right)^{1/2}\right]$ terms in their asymptotic contributions. However, from the asymptotic behaviour
\be
\sqrt{\frac{\omega}{4+\omega}}\left(\frac{\sqrt{4+\omega}-\sqrt{\omega}}{ \sqrt{4+\omega}+\sqrt{\omega}} \right)^2\ \left[\Pi^{(\mu)}(-\omega\ m_\mu^2)\right]
\underset{{\omega\ra\ 0}}{\thicksim}\left(\frac{\alpha}{\pi}\right)\left[ -\frac{1}{30}\omega^{3/2}+\cO(\omega^{2})\right]\,,
\ee
in Eq.~\rf{xandrho1}, there follows that 
\be
\Omega_1 (s)\asymp -\frac{1}{30}\frac{1}{\frac{3}{2}+s}+\cdots \,,
\ee
which  induces a term of $\cO\left[\left(\frac{m_e^2}{m_\mu^2}\right)^{3/2}\right]$ in the asymptotic expansion of $a_{\mu}^{(e\mu)}$. {\it A priori} it could also induce terms of  $\cO\left[\left(\frac{m_e^2}{m_\mu^2}\right)^{3/2}\right]$ in  $a_{\mu}^{(ee\mu)}$. That this is not the case is due to the fact that as we have seen $R_2 (s)$ has zeros at all $s=-1/2 +n$ for $n=-1,-2,-3, \cdots$. However, as we shall later see, $R_3 (s)$ is finite at $s=-3/2$ and  this explains why a term of   $\cO\left[\left(\frac{m_e^2}{m_\mu^2}\right)^{3/2}\right]$ does indeed appear in $a_{\mu}^{(eee\mu)}$. 

\end{itemize}

The  moments $\Omega_0 (s)$ and $R_2 (s)$  fix entirely the evaluation of  the contribution to $a_{\mu}$ from the sixth order Feynman diagram in Fig.~5, which, in our representation, is given by the integral
\begin{equation}\lbl{ee}
	a_{\mu}^{(ee)}  =  \left(\frac{\alpha}{\pi}\right)^3 (-{\bf 1}) \frac{1}{2\pi i}\int\limits_{c_s-i\infty}^{c_s+i\infty}ds \left(\frac{4m_e^2}{m_{\mu}^2} \right)^{-s}\Gamma(s)\Gamma(1-s)\ \Omega_0(s)\ R_{2}(s)\,.
\end{equation}

%%%%%%%%%%%%%%%%%%%%%%%%%%%%%%%%%%%%%%
\begin{figure}[h]

\begin{center}
\includegraphics[width=0.3\textwidth]{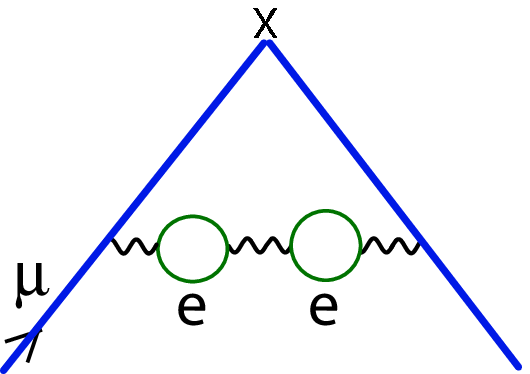}

\end{center}

\vspace*{0.25cm}
{\bf Fig.~5}
{\it\small  Sixth--order Feynman diagram with two lowest order electron--loop vacuum polarization insertions.
}

\end{figure}
%%%%%%%%%%%%%%%%%%%%%%%%%%%%%%%%%%%%%%

\noi
Since we know $\Omega_0(s)$ and $R_{2}(s)$ analytically we can evaluate the full singular series of their product and, therefore, by {\it the converse mapping theorem}, the asymptotic expansion of $a_{\mu}^{(ee)}$ to as high a degree of accuracy as we wish. As an illustration, we give the results for the first few terms:

{\setl
\bea\lbl{eea}
	a_{\mu}^{(ee)}\underset{{\frac{m_{\mu}^2}{m_{e}^2}\ \ra\  \infty}}{\thicksim} &  &  \left(\frac{\alpha}{\pi}\right)^3 \left\{\frac{1}{18}\log^2\frac{m_\mu^2}{m_e^2}  -\frac{25}{54}\log\frac{m_\mu^2}{m_e^2} +\frac{317}{324}+\frac{\pi^2}{27} \right. \nn \\
 & & -\left(\frac{m_e^2}{m_\mu^2}\right)^{1/2}\frac{4}{45}\pi^2  \nn \\
 & & 
 \left.  + \frac{m_e^2}{m_{\mu}^2} \left[-\frac{2}{3}\log^2\frac{m_\mu^2}{m_e^2}+\frac{52}{18}\log\frac{m_\mu^2}{m_e^2}  -4-\frac{4}{9}\pi^2 \right]  +\cO \left[\left(\frac{m_e^2}{m_{\mu}^2}\right)^2 \log^3\frac{m_\mu^2}{m_e^2}\right] \right\}\,,
\eea}

\noi
where the contribution from the leading singularity at $s=0$ is the one in the first line, which agrees with the earlier calculations in refs.~\cite{Ki67,LdeR68}. The exact analytic evaluation of $a_{\mu}^{(ee)}$, as well as of the full sixth--order contribution to $a_{\mu}$ from electron loop insertions, including light--by--light scattering loops, can be found in the papers by Laporta and Remiddi~\cite{La93a,LR93}.

On the other hand, the  moments $\Omega_1 (s)$ and $R_1 (s)$  fix entirely the evaluation of  the contribution to $a_{\mu}$ from the mixed sixth order Feynman diagram in Fig.~6, which, in our representation, is given by the integral
\begin{equation}\lbl{emu}
	a_{\mu}^{(e\mu)}  =  \left(\frac{\alpha}{\pi}\right)^3 (-{\bf 2}) \frac{1}{2\pi i}\int\limits_{c_s-i\infty}^{c_s+i\infty}ds \left(\frac{4m_e^2}{m_{\mu}^2} \right)^{-s}\Gamma(s)\Gamma(1-s)\ \Omega_1(s)\ R_{1}(s)\,.
\end{equation}

%%%%%%%%%%%%%%%%%%%%%%%%%%%%%%%%%%%%%%
\begin{figure}[h]

\begin{center}
\includegraphics[width=0.3\textwidth]{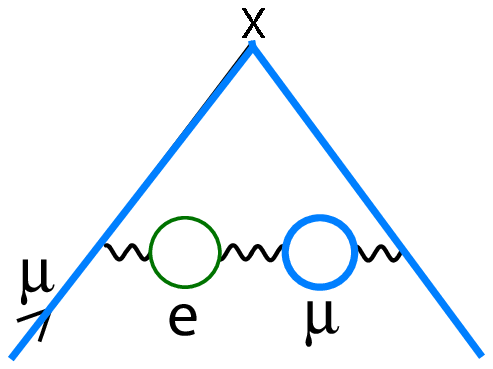}

\end{center}

\vspace*{0.25cm}
{\bf Fig.~6}
{\it\small  Sixth--order Feynman diagram with one electron--loop and one muon loop vacuum polarization insertions.
}

\end{figure}
%%%%%%%%%%%%%%%%%%%%%%%%%%%%%%%%%%%%%%

\noi
Again, since we know $\Omega_1(s)$ and $R_{1}(s)$ we can compute as many terms as we wish of the asymptotic expansion of $a_{\mu}^{(e\mu)}$. We give a few terms below:

{\setl
\bea\lbl{emua}
	a_{\mu}^{(e\mu)}\underset{{\frac{m_{\mu}^2}{m_{e}^2}\ \ra\  \infty}}{\thicksim} &  &  \left(\frac{\alpha}{\pi}\right)^3 \left\{\left(\frac{119}{54}-\frac{2}{9}\pi^2\right)\log\frac{m_\mu^2}{m_e^2} -\frac{61}{162}+\frac{\pi^2}{27} \right. \nn \\
 & & 
 \left.  - \frac{m_e^2}{m_{\mu}^2} \left(\frac{115}{27}+\frac{4}{9}\pi^2\right)   +\cO \left[\left(\frac{m_e^2}{m_{\mu}^2}\right)^2 \log^2\frac{m_\mu^2}{m_e^2}\right]\right\}\,,
\eea}

\noi
in agreement with earlier calculations~\cite{LdeR69,LR93}.

Finally, we observe that once the moments $\Omega_1(s)$ and $R_{2}(s)$ are known, we have all the ingredients to evaluate the eighth--order contribution $a_{\mu}^{(ee\mu)}$ in Eq.~\rf{eemu} which we shall discuss in the next section. With $R_{2}(s)$ known we can also attempt the evaluation of $a_{\mu}^{(ee\tau)}$ in Eq.~\rf{eetau}, which we shall do in Section~VII. 

\begin{itemize}
	\item {\sc The $R_3(s)$ Moment.}

This moment appears in the expressions of $a_{\mu}^{(eee)}$, $a_{\mu}^{(eee\mu)}$ and $a_{\mu}^{(eee\tau)}$ in Eqs.~\rf{eee} \rf{eeemu} and \rf{eeetau} and it is given by the Mellin transform	 
\begin{equation}
		\left(\frac{\alpha}{\pi}\right)^3\ R_3 (s)=\int_{0}^{\infty} d\xi\  \xi^{s-1}\ \rho_3 (\xi)\,,
\end{equation}
of the spectral function in the r.h.s. of Eq.~\rf{dr3}. It gives rise to the  $\delta$--integral

{\setl
\bea\lbl{sigma3}
R_3 (s) & = & \int_0^1 d\delta\ 2\delta (1-\delta^2)^{s-1}\ \left\{ 3\delta\left(\frac{1}{2}-\frac{1}{6}\delta^2\right)\left[\frac{8}{9}-\frac{1}{3}\delta^2+\delta\left(\frac{1}{2}-\frac{1}{6}\delta^2\right) 
\log\frac{1-\delta}{1+\delta}\right]^2 \right.  \nn \\ & & \left.  \hspace*{3.5cm} -\pi^2\left[\delta\left(\frac{1}{2}-\frac{1}{6}\delta^2\right)\right]^3\right\}\,.
\eea}

\noi
Only the terms proportional to $\log^2\frac{1-\delta}{1+\delta}$ require special attention. They give rise to integrals of the type
\be
f(\sigma)=\int_0^1 d\delta (1-\delta^2)^\sigma\log^2\frac{1-\delta}{1+\delta}\quad{\rm with}\quad \sigma=s+n-1\quad\annd\quad n=0,1,2,3,4,5\,. 
\ee
Integrating by parts, we observe that $f(\sigma)$ obeys a simple functional relation
\be
(1+2\sigma)f(\sigma)-2\sigma f(\sigma-1)+\frac{2\sqrt{\pi}}{\sigma}\frac{\Gamma(\sigma)}{\Gamma\left(\frac{1}{2}+\sigma\right)}=0\,,
\ee
from which, and the boundary condition $f(0)=\frac{\pi^2}{3}$,  there follows that
\be\lbl{xifunction}
f(\sigma)=\sqrt{\pi}\ \frac{\Gamma(1+\sigma)}{\Gamma\left(\frac{3}{2}+\sigma\right)}\psi^{(1)}(1+\sigma)\,,\quad\ \rm where   \quad\psi^{(1)}(z)=\frac{d}{dz}\psi(z)  \,.
\ee

One can then do the integral in Eq.~\rf{sigma3}. After some rearrangement so as to exhibit explicitly the singular structure, we get the following expression
\be\lbl{sigma3r}
R_3 (s) =  \frac{\sqrt{\pi}}{864}\frac{\Gamma(s)}{\Gamma\left(\frac{11}{3}+s\right)}
\left[\frac{P_7(s)}{s(1+s)(2+s)}-(1+s)(35+21s+3s^2)\left(27\pi^2-162\ \psi^{(1)}(s)\right) \right]\,,
\ee
with $P_7(s)$ the seventh degree polynomial
\be\lbl{pol7}
P_7(s)=3492 - 8748 s - 26575 s^2 - 9214 s^3 + 18395 s^4 + 17018 s^5 + 
    5120 s^6 + 512 s^7\,.
\ee

\end{itemize}

The function $\psi^{(1)}(z)$  is called the Polygamma function of index one. In general, the Polygamma function of index $n$ is defined as: 
\be
\psi^{(n)}(z)= \frac{d^{n}}{dz^{n}}\psi(z)\,,\quad {\rm with}\quad \psi^{(0)}(z)=\psi(z)\quad {\rm for}\quad n=1,2,3,\dots \,.
\ee
These functions are also related to the Hurwitz function (also called the generalized zeta function)
\be
\zeta(s,z)=\sum_{m=0}^{\infty} (m+z)^{-s}\qquad z\neq 0,-1,-2,\dots\,, \quad {\rm Re} s>1\,,
\ee
as follows 
\be
\psi^{(n)}(z)=(-1)^{n+1} \ n!\  \zeta(n+1,z)\,.
\ee
The Polygamma functions are therefore meromorphic with poles at $z=0,-1,-2,\dots$, with multiplicities $n+1$.
In fact, the Hurwitz function and therefore the Polygamma functions, have the following Mellin representation:
\be
\zeta(s,z)=\frac{1}{\Gamma(s)}\int_0^\infty
dt \ t^{s-1} \ e^{-zt}\ \frac{1}{1-e^{-t}}\qquad \Ree s>1\,,\quad \Ree z>0\,.
\ee

The  moments $\Omega_0 (s)$ and $R_3 (s)$  fix entirely the evaluation of  the contribution to $a_{\mu}$ from the eighth order Feynman diagram in Fig.~1(A), while the moments $\Omega_1 (s)$ and $R_3 (s)$  fix entirely the evaluation of  the contribution to $a_{\mu}$ from the tenth order Feynman diagram in Fig.~2(B). We discuss these results in the next section. With $R_{3}(s)$ known one can also attempt the evaluation of $a_{\mu}^{(eee\tau)}$ in Eq.~\rf{eeetau}, which we shall do in Section~VII.

\begin{itemize}
	\item { \sc The $\Omega_2(s)$ Moment.}
	
It corresponds to the Mellin transform (see Eq.~\rf{xandrho})

\bea\lbl{xandrho2}
	\left(\frac{\alpha}{\pi}\right)^2 \Omega_{2}(s)=\lefteqn{ \int_0^\infty d\omega\ 
 \omega^{s-1} \sqrt{\frac{\omega}{4+\omega}}\left(\frac{\sqrt{4+\omega}-\sqrt{\omega}}{ \sqrt{4+\omega}+\sqrt{\omega}} \right)^2\ \left[\Pi^{(\mu)}(-\omega\ m_\mu^2)\right]^2=}  \nn \\ & &
 \int_0^1 dx\ (1-x)\left(\frac{x^2}{1-x} \right)^s \ 	\left[\Pi^{(\mu)}\left(\frac{-x^2}{1-x}m_\mu^2\right)\right]^2
 \,.
\eea	

\noi
This integral can also be  done, using the expression in  Eq.~\rf{ree} and replacing the logarithms with subtracted logarithms, as in Eq.~\rf{sublog}. One is then left with three types of integrals $\cS^{(j)}_{p}(s,n)$ defined below, where the upper index $j$  refers to the power of the logarithm in the integrand , the lower index $p$ to the degree of the subtracted polynomial and $k\in\mathbb{N}$~:
\be
\cS^{(0)}(s,n)  \equiv  \int_0^1 dx\  x^{2s+n} (1-x)^{1-s} =\frac{\Gamma(2-s)\Gamma(1+n+2s)}{\Gamma(3+n+s)}\,,
\ee 

{\setl
\bea
\cS^{(1)}_{5}(s,n) & \equiv & \int_0^1 dx\  x^{2s+n} (1-x)^{1-s}\left[\log(1-x)-\left(-x-\frac{x^2}{2}-\frac{x^3}{3}-\frac{x^4}{4}-\frac{x^5}{5} \right)\right] \nn \\
& & \nn \\
& = & \frac{1}{60}\Gamma(2-s)\left\{60\frac{\Gamma(2+n+2s)}{\Gamma(4+n+s)}+30
\frac{\Gamma(3+n+2s)}{\Gamma(5+n+s)}       \right.  \nn \\
 & & +20\frac{\Gamma(4+n+2s)}{\Gamma(6+n+s)}+15\frac{\Gamma(5+n+2s)}{\Gamma(7+n+s)}+
 12\frac{\Gamma(6+n+2s)}{\Gamma(8+n+s)}\nn \\
 & & \left. +60\frac{\Gamma(1+n+2s)}{\Gamma(3+n+s)}\left(\rH_{1-s} -\rH_{2+n+s} \right)\right\}\,, 
 \eea}
 
\noi
and

{\setl
\bea 
    \cS^{(2)}_{6}(s,n) & \equiv & \int_0^1 dx\  x^{2s+n} (1-x)^{1-s}\left[\log^2(1-x)-\left(x^2+x^3 +\frac{11x^4}{12}+\frac{5x^5}{6}+\frac{137x^6}{180}\right)\right]\nn \\
       & & \nn \\
    & = & -\frac{1}{180}\Gamma(2-s)\left\{-180\frac{\Gamma(3+n+2s)}{\Gamma(5+n+s)}-180
\frac{\Gamma(4+n+2s)}{\Gamma(6+n+s)}       \right.  \nn \\
 & & -165\frac{\Gamma(5+n+2s)}{\Gamma(7+n+s)}-150\frac{\Gamma(6+n+2s)}{\Gamma(8+n+s)}-
 137\frac{\Gamma(7+n+2s)}{\Gamma(9+n+s)}\nn \\
 & & +\frac{180}{\Gamma(3+n+s)}\left[\Gamma(1+n+2s)\left(\psi^2(2-s)-2\psi(2-s)\psi(3+n+s) +\right. \right. \nn \\
 & & \left. \left.\left. + \psi^2(3+n+s)+ \psi^{(1)}(2-s) -\psi^{(1)}(3+n+s)\right)\right]\right\}\,. 
\eea}

\noi
The analytic result for $\Omega_{2}(s)$ can then  be expressed in terms of these three integrals as follows:

{\setl
\bea\lbl{omega2}
\Omega_{2}(s) & = & 
-\frac{88}{135}\ \cS^{(0)}(s,1)+\frac{217}{135}\ \cS^{(0)}(s,2)-\frac{3}{5}\ \cS^{(0)}(s,3)
-\frac{347}{540}\ \cS^{(0)}(s,4)\nn \\
 & & +\frac{1}{6}\ \cS^{(0)}(s,5)+\frac{137}{1620}\ \cS^{(0)}(s,6)\nn \\
 & & \nn \\
  & & 
 -\frac{10}{27}\ \cS^{(1)}_{5}(s,0)-\frac{8}{9}\ \cS^{(1)}_{5}(s,-1)+\frac{28}{9}\ \cS^{(1)}_{5}(s,-2)
+\frac{104}{27}\ \cS^{(1)}_{5}(s,-3)\nn \\
& & -\frac{80}{9}\ \cS^{(1)}_{5}(s,-4)+\frac{32}{9}\ \cS^{(1)}_{5}(s,-5)\nn \\
& & \nn \\
& & +\frac{1}{9}\ \cS^{(2)}_{6}(s,0)-\frac{4}{3}\ \cS^{(2)}_{6}(s,-2)+\frac{8}{9}\ \cS^{(2)}_{6}(s,-3)
+4\ \cS^{(2)}_{6}(s,-4)\nn \\
 & & -\frac{16}{3}\ \cS^{(2)}_{6}(s,-5)+\frac{16}{9}\ \cS^{(2)}_{6}(s,-6)\,.
\eea}

In particular, the value of $\Omega_{2}(s)$ at $s=0$ fixes the contribution to the muon anomaly from the Feynman diagram in Fig.~7, i.e.

%%%%%%%%%%%%%%%%%%%%%%%%%%%%%%%%%%%%%%
\begin{figure}[h]

\begin{center}
\includegraphics[width=0.3\textwidth]{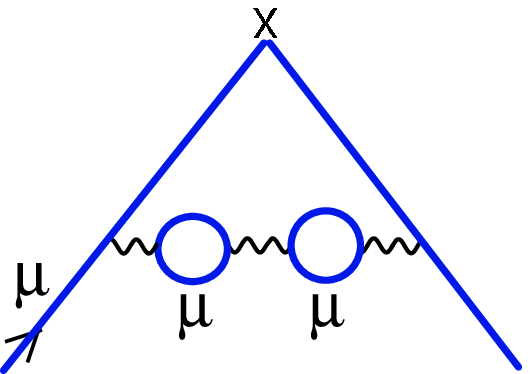}

\end{center}

\vspace*{0.25cm}
{\bf Fig.~7}
{\it\small  Sixth--order Feynman diagram with two lowest order muon--loop vacuum polarization insertions.
}

\end{figure}
%%%%%%%%%%%%%%%%%%%%%%%%%%%%%%%%%%%%%%
\be
a_{\mu}^{(\mu\mu)}=\left( \frac{\alpha}{\pi}\right)^3 \Omega_{2}(0)=\left( \frac{\alpha}{\pi}\right)^3 \left[ -\frac{943}{324}-\frac{4}{135}\pi^2+\frac{8}{3}\zeta(3)\right]\,.
\ee

\end{itemize}

Knowing $\Omega_{2}(s)$ and $R_1 (s)$ we have all the ingredients to evaluate $a_{\mu}^{(e\mu\mu)}$ in Fig.~1C and knowing $\Omega_{2}(s)$ and $R_2 (s)$ we can also evaluate $a_{\mu}^{(ee\mu\mu)}$ in Fig.~2C. We discuss these calculations in the next section.

\begin{itemize}
	\item {\sc The $R_4(s)$ Moment.}

This moment appears in the expression of $a_{\mu}^{(eeee)}$ and it is given by the Mellin transform	 
\begin{equation}
		\left(\frac{\alpha}{\pi}\right)^4\ R_4 (s)=\int_{0}^{\infty} d\xi\  \xi^{s-1}\ \rho_4 (\xi)\,,
\end{equation}
of the spectral function in the r.h.s. of Eq.~\rf{dr4}. It gives rise to the  $\delta$--integral

{\setl
\bea
R_4 (s) & = & \int_0^1 d\delta\ 2\delta (1-\delta^2)^{s-1}\ \left\{ 4\delta\left(\frac{1}{2}-\frac{1}{6}\delta^2\right)\left[\frac{8}{9}-\frac{1}{3}\delta^2+\delta\left(\frac{1}{2}-\frac{1}{6}\delta^2\right) 
\log\frac{1-\delta}{1+\delta}\right]^3 \right.  \nn \\ & & \left.   -4\pi^2
\left[\frac{8}{9}-\frac{1}{3}\delta^2+\delta\left(\frac{1}{2}-\frac{1}{6}\delta^2\right) 
\log\frac{1-\delta}{1+\delta}\right]\left[\delta\left(\frac{1}{2}-\frac{1}{6}\delta^2\right)\right]^3\right\}\,.
\eea}

\noi
Here only the integrals proportional to $\log^3\frac{1-\delta}{1+\delta}$ are new with respect to the integrals which already appeared in the evaluation of $R_2 (s)$. By integration by parts, they can be reduced to integrals proportional to $\log^2\frac{1-\delta}{1+\delta}$ and hence to the function $f(s)$ in Eq.~\rf{xifunction}. In terms of $f(s)$ and the auxiliary function
\be
g(\sigma)=2\sqrt{\pi}\frac{\Gamma(\sigma)}{\Gamma\left(\sigma+\frac{1}{2}\right)}=2 B\left(\frac{1}{2},\sigma\right)\,,
\ee
we then find:

{\setl
\bea\lbl{sigma4}
 R_4 (s) & =  & \left(-\frac{250}{2187}+\frac{10 }{243}\pi^2 \right) g(s)-
 \left(\frac{325}{2187}-\frac{\pi^2}{243} \right) g(s+1)+
 \left(\frac{80}{2187}-\frac{31}{486} \pi^2\right) g(s+2)\nn \\ 
 & & +
 \left(\frac{34}{243}-\frac{25}{972}\pi^2 \right) g(s+3)+
 \left(\frac{2}{27}+\frac{23}{972}\pi^2 \right) g(s+4)\nn \\ 
 & & +
 \left(\frac{1}{81}+\frac{17}{972}\pi^2 \right) g(s+5)+
 \frac{\pi^2}{324}  g(s+6)\nn \\
  & & \nn \\
  & & +\left(\frac{50}{243}-\frac{2}{81}\pi^2 \right)\frac{g(s)}{s}+
 \frac{20}{81}\frac{g(s+1)}{s+1}-\left(\frac{13}{162}-\frac{\pi^2}{27} \right)\frac{g(s+2)}{s+2}\nn \\
  & & -
 \left(\frac{115}{486}-\frac{\pi^2}{81} \right)\frac{g(s+3)}{s+3}
 -\left(\frac{19}{162}+\frac{\pi^2}{72} \right)\frac{g(s+4)}{s+4}\nn \\
  & & -
 \left(\frac{1}{54}+\frac{\pi^2}{108} \right)\frac{g(s+5)}{s+5}-
\frac{\pi^2}{648}\frac{g(s+6)}{s+6}  \nn \\
 & & \nn \\
  & & +\frac{1}{81}\left[-40\ f(-1+s)-4\ f(s)+62\ f(1+s)+25\ f(2+s)\right.\nn \\
   & & \left. -23\ f(3+s)
  -17\ f(4+s)-3\ f(5+s) \right]\nn \\
  & & \nn \\
   & & +\frac{1}{54}\left[16\frac{f(-1+s)}{s}-24\frac{f(1+s)}{s+2}-8
   \frac{f(2+s)}{s+3}\right.\nn \\
   & & \left. +9 \frac{f(3+s)}{s+4}+6\frac{f(4+s)}{s+5}+\frac{f(5+s)}{s+6}\right]\,.
 \eea}
 
 \noi

\end{itemize}
Knowing explicitly the function $R_4 (s)$ will allow us to perform the evaluation of $a_{\mu}^{(eeee)}$ which we do in the next section.  

\begin{itemize}
	\item { \sc The $\Omega_3(s)$ Moment.}
	
It corresponds to the Mellin transform (see Eq.~\rf{xandrho})

\bea\lbl{xandrho3}
	\left(\frac{\alpha}{\pi}\right)^2 \Omega_{3}(s)=\lefteqn{ \int_0^\infty d\omega\ 
 \omega^{s-1} \sqrt{\frac{\omega}{4+\omega}}\left(\frac{\sqrt{4+\omega}-\sqrt{\omega}}{ \sqrt{4+\omega}+\sqrt{\omega}} \right)^2\ \left[\Pi^{(\mu)}(-\omega\ m_\mu^2)\right]^3=}  \nn \\ & &
 \int_0^1 dx\ (1-x)\left(\frac{x^2}{1-x} \right)^s \ 	\left[\Pi^{(\mu)}\left(\frac{-x^2}{1-x}m_\mu^2\right)\right]^3
 \,.
\eea	

\noi
As in the evaluation of the integral in Eq.~\rf{xandrho2}, the first step consists in replacing the logarithms and their powers by subtracted logarithms and subtracted power logarithms. One is then left with  integrals of the type $\cS^{(j)}_{p}(s,n)$ like those already introduced in the evaluation of $\Omega_{2}(s)$. The new ones are:

{\setl
\bea
\cS^{(1)}_{7}(s,n) & \equiv & \int_0^1 dx\  x^{2s+n} (1-x)^{1-s}\left[\log(1-x)-\left(-x-\frac{x^2}{2}-\frac{x^3}{3}-\frac{x^4}{4}-\frac{x^5}{5}-\frac{x^6}{6} -\frac{x^7}{7}  \right)\right] \nn \\
& & \nn \\
& = & \cS^{(1)}_5(s,n)+\frac{1}{42}\Gamma(2-s)\left[7\ \frac{\Gamma(7+n+2s)}{\Gamma(9+n+s)} +6\ \frac{\Gamma(8+n+2s)}{\Gamma(10+n+s)}\right]\,, 
 \eea}
 
 \noi

{\setl
\bea 
    \cS^{(2)}_{8}(s,n) & \equiv & \int_0^1 dx\  x^{2s+n} (1-x)^{1-s}\left[\log^2(1-x)\ -\right.\nn \\ 
    & &\left. \left(x^2+x^3 +\frac{11x^4}{12}+\frac{5x^5}{6}+\frac{137x^6}{180}+\frac{7x^7}{10}+
    \frac{363x^8}{560}\right)\right]\nn \\
       & & \nn \\
    & = & \cS^{(2)}_{6}(s,n)-\frac{1}{5040}\Gamma(2-s)\!\left[3528 \frac{\Gamma(8+n+2s)}{\Gamma(10+n+s)} + 3267 \frac{\Gamma(9+n+2s)}{\Gamma(11+n+s)}\right]\,,
\eea}

\noi

and

{\setl
\bea 
    \cS^{(3)}_{9}(s,n) & \equiv & \int_0^1 dx\  x^{2s+n} (1-x)^{1-s}\left[\log^3(1-x)\ -\right.\nn \\ 
    & &\left. -\left(-x^3 -\frac{3x^4}{2}-\frac{7x^5}{4}-\frac{15x^6}{8}-\frac{29x^7}{15}-
    \frac{469x^8}{240}-\frac{29531x^9}{240}\right)\right]\nn \\
       & & \nn \\
    & = &   \frac{1}{15120}\Gamma(2-s)\left\{15120\frac{\Gamma(4+n+2s)}{\Gamma(6+n+s)}+22680
\frac{\Gamma(5+n+2s)}{\Gamma(7+n+s)}+26460\frac{\Gamma(6+n+2s)}{\Gamma(6+n+s)}       \right.  \nn \\
 & & +28350\frac{\Gamma(7+n+2s)}{\Gamma(9+n+s)}+
 29232\frac{\Gamma(8+n+2s)}{\Gamma(10+n+s)}\nn \\
  & & +29547\frac{\Gamma(9+n+2s)}{\Gamma(11+n+s)}
 +29531\frac{\Gamma(10+n+2s)}{\Gamma(12+n+s)}\nn \\
 & & +\frac{15120}{\Gamma(3+n+s)}\left[\Gamma(1+n+2s)\left(\psi^3(2-s)-3\psi(2-s)^2
 \psi(3+n+s)- \psi^3(3+n+s) \right. \right. \nn \\
 & &  -3\psi(3+n+s)\left[ \psi^{(1)}(2-s) -\psi^{(1)}(3+n+s)\right]\nn \\
  & &  +3\psi(2-s)\left[\psi^2 (3+n+s)+\psi^{(1)}(2-s)-
  \psi^{(1)}(3+n+s) \right]\nn \\
   & &\left.\left.\left. +\psi^{(2)}(2-s)-\psi^{(2)}(3+n+s)\right)\right]\right\}\,. 
\eea}

\noi

The analytic result for $\Omega_{3}(s)$ can then  be expressed in terms of the $\cS^{(j)}_p (s,n)$  integrals as follows:

{\setl
\bea\lbl{omega3}
\Omega_{3}(s) & = & 
-\frac{482}{405}\ \cS^{(0)}(s,1)+\frac{3007}{567}\ \cS^{(0)}(s,2)-\frac{125281}{17010}\ \cS^{(0)}(s,3)
-\frac{16729}{11340}\ \cS^{(0)}(s,4)+\frac{8329}{2268}\ \cS^{(0)}(s,5)\nn \\
 & & -\frac{11726}{8505}\ \cS^{(0)}(s,6)-\frac{1937}{2520}\ \cS^{(0)}(s,7)+\frac{1091}{5670}\ \cS^{(0)}(s,8)
 +\frac{29531}{408240}\ \cS^{(0)}(s,9)\nn \\
 & & \nn \\
  & & 
 -\frac{25}{81}\ \cS^{(1)}_{7}(s,0)-\frac{40}{27}\ \cS^{(1)}_{7}(s,-1)+\frac{14}{9}\ \cS^{(1)}_{7}(s,-2)
+\frac{908}{81}\ \cS^{(1)}_{7}(s,-3)\nn \\
& & -\frac{160}{27}\ \cS^{(1)}_{7}(s,-4)-\frac{608}{27}\ \cS^{(1)}_{7}(s,-5)
+\frac{224}{9}\ \cS^{(1)}_{7}(s,-6)-\frac{64}{9}\ \cS^{(1)}_{7}(s,-7)
\nn \\
& & \nn \\
& & +\frac{5}{27}\ \cS^{(2)}_{8}(s,0)+\frac{4}{9}\ \cS^{(2)}_{8}(s,-1)
-\frac{8}{3}\ \cS^{(2)}_{8}(s,-2)-\frac{104}{27}\ \cS^{(2)}_{8}(s,-3)
+\frac{140}{9}\ \cS^{(2)}_{8}(s,-4)\nn \\
 & & 
+\frac{32}{9}\ \cS^{(2)}_{8}(s,-5)-\frac{928}{27}\ \cS^{(2)}_{8}(s,-6)+
\frac{256}{9}\ \cS^{(2)}_{8}(s,-7)-\frac{64}{9}\ \cS^{(2)}_{8}(s,-8)
\nn \\
& & \nn \\
 & & 
 -\frac{1}{27}\ \cS^{(3)}_{9}(s,0)+\frac{2}{3}\ \cS^{(3)}_{9}(s,-2)-\frac{4}{9}\ \cS^{(3)}_{9}(s,-3)-4\ \cS^{(3)}_{9}(s,-4)+\frac{16}{3}\ \cS^{(3)}_{9}(s,-5)\nn \\
  & & +\frac{56}{9}\ \cS^{(3)}(s,-6)-16\ \cS^{(3)}(s,-7)+\frac{32}{3}\ \cS^{(3)}_{9}(s,-8)-\frac{64}{27}\ \cS^{(3)}_{9}(s,-9)\,.
\eea}

In particular, the value of $\Omega_{3}(s)$ at $s=0$ fixes the contribution to the muon anomaly from the Feynman diagram in Fig.~8, i.e.

%%%%%%%%%%%%%%%%%%%%%%%%%%%%%%%%%%%%%%
\begin{figure}[h]

\begin{center}
\includegraphics[width=0.4\textwidth]{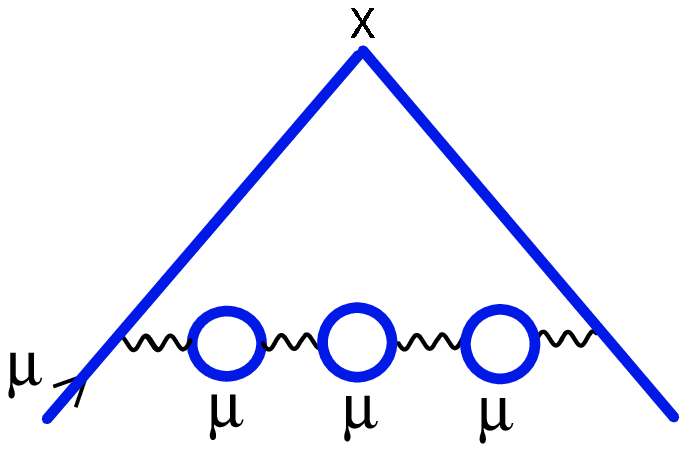}

\end{center}

\vspace*{0.25cm}
{\bf Fig.~8}
{\it\small  Eighth--order Feynman diagram with three lowest order muon--loop vacuum polarization insertions.
}

\end{figure}
%%%%%%%%%%%%%%%%%%%%%%%%%%%%%%%%%%%%%%
\be
a_{\mu}^{(\mu\mu\mu)}=\left( \frac{\alpha}{\pi}\right)^4 \Omega_{3}(0)=
\left( \frac{\alpha}{\pi}\right)^4 \left[
-\frac{151849}{40824}+\frac{2}{45}\pi^4-\frac{32}{63}\zeta(3) \right]\,.
\ee

\end{itemize}

Knowing $\Omega_{3}(s)$ and $R_1 (s)$ we have all the ingredients to evaluate $a_{\mu}^{(e\mu\mu\mu)}$ in Fig.~2D. We discuss these calculations in the next section.

%%%%%%%%%%%%%%%%%%%%%%%%%%%%%%%%%%%%%%%%%%%%%%%%%%%%%%%%%%%%%%%%%%%%%%%%
%%%%%%%%%%%%%%%%%%%%%%%%%%%%%%%%%%%%%%%%%%%%%%%%%%%%%%%%%%%%%%%%%%%%%%%%
%%%%%%%%%%%%%%%%%%%%%%%%%%%%%%%%%%%%%%%%%%%%%%%%%%%%%%%%%%%%%%%%%%%%%%%%
%%%%%%%%%%%%%%%%%%%%%%%%%%%%%%%%%%%%%%%%%%%%%%%%%%%%%%%%%%%%%%%%%%%%%%%%
%%%%%%%%%%%%%%%%%%%%%%%%%%%%%%%%%%%%%%%%%%%%%%%%%%%%%%%%%%%%%%%%%%%%%%%%
\section{\small Eighth Order Results from Electron and Muon Vacuum Polarization Loops}
\setcounter{equation}{0}
\def\theequation{\arabic{section}.\arabic{equation}}

\noi
We have now all the ingredients to proceed to the calculation of the eighth order contributions illustrated by the Feynman diagrams in Fig.~1. 

\begin{itemize}
	\item {\sc Three Electron Loops}, Fig.~1(A) [one diagram]:
	
	We recall the expression in Eq.~\rf{eee}
	
\be\lbl{eeeR}
	a_{\mu}^{(eee)}  =  \left(\frac{\alpha}{\pi}\right)^4 \frac{1}{2\pi i}\int\limits_{c_s-i\infty}^{c_s+i\infty}ds \left(\frac{4m_e^2}{m_{\mu}^2} \right)^{-s}\Gamma(s)\Gamma(1-s)\ \Omega_{0}(s)\ R_{3}(s)\,.
\ee
The {\it converse mapping theorem} relates the asymptotic behaviour of  $a_{\mu}^{(eee)}$ as a function of the small mass ratio ${4m_e^2}/{m_{\mu}^2}$ to the singularities of the integrand in this equation as a function of the Mellin $s$--complex variable.
Using the explicit expressions for $\Omega_{0}(s)$ and $R_{3}(s)$ given in Eqs.~\rf{omega0} and \rf{sigma3} we can proceed to the calculation of the singular series expansion of the integrand in question. The relevant singularities are those in the left--hand side of the Mellin $s$--plane. They occur as multipoles at $s=0,-1,-2,-3,\dots$, because of the factors $\Gamma(s)$ and $R_{3}(s)$; and as single poles  at $s=-1/2,-3/2,-5/2,\dots$ because of the factor $\Omega_{0}(s)$. The leading singularity at $s\ra 0$ has a quadruple pole, a triple pole, a double pole and a single pole. Their residues govern the successive terms of $\cO(\log^3 \frac{m_{\mu}^2}{m_{e}^2})$, of $\cO(\log^2 \frac{m_{\mu}^2}{m_{e}^2})$, of $\cO(\log \frac{m_{\mu}^2}{m_{e}^2})$ and of $\cO({\rm Cte.})$ in the leading  contributions to $a_{\mu}^{(eee)}$ for $\frac{m_{\mu}^2}{m_{e}^2}$ large:

{\setl
\bea
a_{\mu}^{(eee)} &  \underset{{[s\ \ra\  0]}}{\thicksim} & \left(\frac{\alpha}{\pi}\right)^4\left[  \frac{1}{54} \log^3 \frac{m_{\mu}^2}{m_{e}^2}-\frac{25}{108} \log^2 \frac{m_{\mu}^2}{m_{e}^2}+\left(\frac{317}{324}+\frac{\pi^2}{27}\right) \log \frac{m_{\mu}^2}{m_{e}^2} \right.\nn \\
 & & \left. -\frac{8609}{5832}-\frac{25}{162}\pi^2-\frac{2}{9}\zeta(3)\right]\,. 
\eea}

\noi
The next to leading singularity at $s\ra -1/2$ is a simple pole, induced by $\Omega_{0}(-1/2)$, which governs the contribution of $\cO\left(\frac{m_{e}}{m_{\mu}}\right)$:
\be
a_{\mu}^{(eee)} \underset{{\left[s\ \ra\   -\frac{1}{2}\right]}}{\thicksim}\left(\frac{\alpha}{\pi}\right)^4  \frac{m_{e}}{m_{\mu}}\ \frac{101}{1536}\pi^4\,.
\ee
The next singularity at $s\ \ra\ -1$ in the Mellin plane is again a multipole type singularity which, by the inverse mapping theorem, governs the terms

{\setl
\bea
a_{\mu}^{(eee)}  &  \underset{{[s\ \ra\   -1]}}{\thicksim} & \left(\frac{\alpha}{\pi}\right)^4   \frac{m_{e}^2}{m_{\mu}^2}\left[
-\frac{2}{9}\log^3 \frac{m_{\mu}^2}{m_{e}^2}+\frac{13}{9}\log^2 \frac{m_{\mu}^2}{m_{e}^2}-\left( \frac{152}{27}+\frac{4}{9}\pi^2\right)\log \frac{m_{\mu}^2}{m_{e}^2} \right. \nn \\
& & \left. +\frac{967}{315}+\frac{26}{27}\pi^2+\frac{136}{35}\zeta(3)\right]\,.
\eea}

\noi
There is no problem in evaluating as many terms as we wish. In the Appendix we give the result of the asymptotic contribution to $a_{\mu}^{(eee)}$ up to terms of $\cO\left[\left(\frac{\alpha}{\pi}\right)^4 \left(\frac{m_{e}^2}{m_{\mu}^2}\right)^{5/2}\right]$.  Because of the present experimental error in the $\frac{m_{\mu}}{m_{e}}$ ratio, there is no need to go beyond this approximation, but we have checked that including higher order terms  up to $\cO\left[\left(\frac{\alpha}{\pi}\right)^4 \left(\frac{m_{e}^2}{m_{\mu}^2}\right)^{3}\right]$ does not change our final numerical result in Table~1 in the Appendix. The terms up to $\cO\left[\left(\frac{\alpha}{\pi}\right)^4 \left(\frac{m_{e}^2}{m_{\mu}^2}\right)^{3/2}\right]$ agree with those obtained by Laporta~\cite{La93}  using a very different method. The  agreement with the numerical determination by Kinoshita and Nio~\cite{KN04} is quite remarkable, although our result is of course more precise.

	\item {\sc Two Electron Loops and One Muon Loop}, Fig.~1(B) [three diagrams]:
	
The corresponding expression is the one in Eq.~\rf{eemu}	
\be
	a_{\mu}^{(ee\mu)}  =  \left(\frac{\alpha}{\pi}\right)^4  {\bf 3}\ \frac{1}{2\pi i}\int\limits_{c_s-i\infty}^{c_s+i\infty}ds \left(\frac{4m_e^2}{m_{\mu}^2} \right)^{-s}\Gamma(s)\Gamma(1-s)\ \Omega_1 (s)\ R_2 (s)\,,
\ee
with $\Omega_1 (s)$ and $R_2 (s)$ given in Eqs.~\rf{omega1} and \rf{sigma2}. The relevant singularities in the Mellin $s$--plane occur at $s=0,-1,-2,-3,\cdots$ as multipoles. The  singularity at $s=0$ governs the terms

{\setl
\bea\lbl{eemu0}
a_{\mu}^{(ee\mu)} &  \underset{{[s\ \ra\  0]}}{\thicksim} & \left(\frac{\alpha}{\pi}\right)^4\left[  \left(\frac{119}{108}-\frac{\pi^2}{9}\right) \log^2 \frac{m_{\mu}^2}{m_{e}^2}-\left(\frac{61}{162}-\frac{\pi^2}{27} \right)\log \frac{m_{\mu}^2}{m_{e}^2}\right. \nn \\
& & \left.
+\frac{7627}{1944}+\frac{13}{27}\pi^2-\frac{4}{45}\pi^4 \right]\,, 
\eea}

\noi
in the asymptotic expansion. The next--to --leading singularity is at $s=-1$ and it is responsible for the higher order terms

{\setl
\bea\lbl{eemu1}
a_{\mu}^{(ee\mu)} &  \underset{{[s\ \ra\  -1]}}{\thicksim} & \left(\frac{\alpha}{\pi}\right)^4 \left(\frac{m_{e}^2} {m_{\mu}^2}\right) 
\left[ \left(-\frac{115}{27}+\frac{4}{9}\pi^2 \right) \log\frac{m_{\mu}^2}{m_{e}^2}+
\frac{227}{18}-\frac{4}{3}\pi^2\right]\,.
\eea}

\noi
The analytic results in Eqs.~\rf{eemu0} and \rf{eemu1} agree with those given by Laporta in ref.~\cite{La93}. 
As discussed previously, there is no singularity at $s=-1/2$ because of the presence of a muon loop self--energy (see the discussion around Eq.~\rf{selfmu0} ). There could be {\it a priori}  a singularity at $s=-3/2$ because it appears in the singular expansion of $\Omega_1(s)$; however, it is screened by the fact that $R_2(s)$ has a zero at $s=-3/2$.
 In the Appendix we give the result for $a_{\mu}^{(ee\mu)}$ including terms of $\cO\left[\left(\frac{\alpha}{\pi}\right)^4 \left(\frac{m_{e}^2}{m_{\mu}^2}\right)^{2}\right]$. Taking into account higher order terms brings in contributions which numerically are of the order of the error induced by the present value of the $\frac{m_{\mu}^2}{m_{e}^2}$ ratio in the previous terms.  Numerically, our result is more precise than  the one obtained by Kinoshita and Nio~\cite{KN04}, though it agrees very well within their quoted error. 

	\item {\sc One Electron Loop and Two Muon Loops}, Fig.~1(C) [three diagrams]:

The corresponding expression is the one in Eq.~\rf{emumu}
\be	a_{\mu}^{(e\mu\mu)}  =   \left(\frac{\alpha}{\pi}\right)^4 {\bf 3}\ \frac{1}{2\pi i}\int\limits_{c_s-i\infty}^{c_s+i\infty}ds \left(\frac{4m_e^2}{m_{\mu}^2} \right)^{-s}\Gamma(s)\Gamma(1-s)\ \Omega_2 (s)\ R_1 (s) \,,
\ee 
with $\Omega_2 (s)$ and $R_1 (s)$ given in Eqs.~\rf{omega2} and \rf{sigma1}.  The relevant singularities in the Mellin $s$--plane occur at $s=0,-1,-2,-3,\cdots$.  The  singularity at $s=0$ is a double pole and governs the terms  

{\setl
\bea\lbl{emumu0}
a_{\mu}^{(e\mu\mu)} &  \underset{{[s\ \ra\  0]}}{\thicksim} & \left(\frac{\alpha}{\pi}\right)^4\left[\left(-\frac{943}{324}-\frac{4}{135}\pi^2+\frac{8}{3}\zeta(3) \right)\log \frac{m_{\mu}^2}{m_{e}^2}\right. \nn \\
 & & 
\left. +\frac{57899}{9720}-\frac{5383}{4050}\pi^2+\frac{2}{27}\pi^4-\frac{2}{45}\zeta(3) \right]\,. 
\eea}

The next--to--leading singularity at $s=-1$ is a single pole which governs the next--to--leading term

{\setl
\bea\lbl{emumu1}
a_{\mu}^{(e\mu\mu)} &  \underset{{[s\ \ra\  -1]}}{\thicksim} & \left(\frac{\alpha}{\pi}\right)^4\ \left(\frac{m_{e}^2}{m_{\mu}^2}\right)\left[\frac{458}{81}-\frac{26}{105}\pi^2-\frac{8}{3}\zeta(3) \right]\,; 
\eea}

\noi 
and so on.  The results in Eqs.~\rf{emumu0} and \rf{emumu1} agree with those of Laporta~\cite{La93}. In the appendix we have also included terms up to  $\cO\left[\left(\frac{\alpha}{\pi}\right)^4 \left(\frac{m_{e}^2}{m_{\mu}^2}\right)^{2}\right]$. These are the terms one needs to fix  $a_{\mu}^{(e\mu\mu)}$ to the accuracy  required by the present knowledge of the $\frac{m_{e}^2}{m_{\mu}^2}$ mass ratio.  It is not surprising that  this analytic evaluation allows for a more accurate determination of this term than the numerical estimate by Kinoshita and Nio~\cite{KN04}, though it agrees with it within their given error.

\end{itemize}

%%%%%%%%%%%%%%%%%%%%%%%%%%%%%%%%%%%%%%%%%%%%%%%%%%%%%%%%%%%%%%%%%%%%%%%%
%%%%%%%%%%%%%%%%%%%%%%%%%%%%%%%%%%%%%%%%%%%%%%%%%%%%%%%%%%%%%%%%%%%%%%%%
%%%%%%%%%%%%%%%%%%%%%%%%%%%%%%%%%%%%%%%%%%%%%%%%%%%%%%%%%%%%%%%%%%%%%%%%
%%%%%%%%%%%%%%%%%%%%%%%%%%%%%%%%%%%%%%%%%%%%%%%%%%%%%%%%%%%%%%%%%%%%%%%%
%%%%%%%%%%%%%%%%%%%%%%%%%%%%%%%%%%%%%%%%%%%%%%%%%%%%%%%%%%%%%%%%%%%%%%%%
%%%%%%%%%%%%%%%%%%%%%%%%%%%%%%%%%%%%%%%%%%%%%%%%%%%%%%%%%%%%%%%%%%%%%%%%
\section{\small Tenth Order Results from Electron and Muon Vacuum Polarization Loops}
\setcounter{equation}{0}
\def\theequation{\arabic{section}.\arabic{equation}}

\noi
We also have all the ingredients to proceed to the calculation of the tenth order contributions illustrated by the Feynman diagrams in Fig.~2. 

\begin{itemize}
	\item {\sc Four Electron Loops}, Fig.~2(A) [one diagram]:
	
	We recall the expression in Eq.~\rf{eeee}
	\be
	a_{\mu}^{(eeee)}  =  \left(\frac{\alpha}{\pi}\right)^5\ (-{\bf 1}) \frac{1}{2\pi i}\int\limits_{c_s-i\infty}^{c_s+i\infty}ds \left(\frac{4m_e^2}{m_{\mu}^2} \right)^{-s}\Gamma(s)\Gamma(1-s)\ \Omega_{0}(s)\ R_4 (s)\,,
\ee
with $\Omega_{0}(s)$ and  $R_4 (s)$ given in Eqs.~\rf{omega0} and \rf{sigma4}. The {\it converse mapping theorem} relates the asymptotic behaviour of  $a_{\mu}^{(eee)}$ as a function of the small mass ratio ${4m_e^2}/{m_{\mu}^2}$, to the singular series expansion of the integrand in this equation. 
The relevant singularities are those in the left--hand side of the Mellin $s$--plane. They occur as multipoles at $s=0,-1,-2,-3,\dots$, because of the factors $\Gamma(s)$ and $R_{4}(s)$; and as single poles  at $s=-1/2,-3/2,-5/2,\dots$ because of the factor $\Omega_{0}(s)$. The multipoles at $s\ra 0$ govern the asymptotic contributions:

{\setl
\bea\lbl{eeee0}
a_{\mu}^{(eeee)} &  \underset{{[s\ \ra\  0]}}{\thicksim} & \left(\frac{\alpha}{\pi}\right)^5\left\{\frac{1}{162}\log^4 \frac{m_{\mu}^2}{m_{e}^2}-\frac{25}{243}\log^3 \frac{m_{\mu}^2}{m_{e}^2}+
\left(\frac{317}{486}+\frac{2}{81}\pi^2 \right) \log^2 \frac{m_{\mu}^2}{m_{e}^2}\right. \nn \\
 & & 
-\left(\frac{8609}{4374}+\frac{50}{243}\pi^2+\frac{8}{27}\zeta(3) \right)\log \frac{m_{\mu}^2}{m_{e}^2} \nn \\
& & \left.  +\frac{64613}{26244}+\frac{317}{729}\pi^2+\frac{2}{135}\pi^4+\frac{100}{81}\zeta(3) \right\}\,. 
\eea}

\noi
There is only a simple pole at $s=-1/2$. It governs the next--to--leading asymptotic contribution

{\setl
\bea\lbl{eeee1}
a_{\mu}^{(eeee)} &  \underset{{[s\ \ra\  -1/2]}}{\thicksim} & \left(\frac{\alpha}{\pi}\right)^5 \left(\frac{m_{e}^2}{m_{\mu}^2} \right)^{1/2}\left[-\frac{18203}{374220}\pi^4\right]\,. 
\eea}

\noi
The contributions in Eqs.~\rf{eeee0} and \rf{eeee1} agree with those given by Laporta in ref.~\cite{La94}. We have calculated 
further contributions, up to the required accuracy governed by the present knowledge of the $\frac{m_{e}^2}{m_{\mu}^2}$ ratio. They are  given in the Appendix. Numerically, although our result is much more precise than the one given by Kinoshita and Nio~\cite{KN04}, it agrees very well with it within their quoted errors.

	\item {\sc Three Electron Loops and One Muon Loop}, Fig.~2(B) [four diagrams]:

	We recall the expression in Eq.~\rf{eeemu}	
\be
	a_{\mu}^{(eee\mu)}  =  \left(\frac{\alpha}{\pi}\right)^5  (-{\bf 4})\ \frac{1}{2\pi i}\int\limits_{c_s-i\infty}^{c_s+i\infty}ds \left(\frac{4m_e^2}{m_{\mu}^2} \right)^{-s}\Gamma(s)\Gamma(1-s)\ \Omega_1 (s)\ R_3 (s)\,,
\ee
with $\Omega_1 (s)$ and  $R_3 (s)$ given in Eqs.~\rf{omega1} and \rf{sigma3}. 
The relevant singularities here occur as multipoles at $s=0,-1,-2,-3,\dots$, because of the factors $\Gamma(s)$ and $R_{3}(s)$; and as single poles  at $s=-3/2,-5/2,\dots$ because of the factor $\Omega_{1}(s)$. The multipoles at $s\ra 0$ govern the asymptotic contributions:

{\setl
\bea
a_{\mu}^{(eee\mu)} & \underset{{[s\ \ra\  0]}}{\thicksim} & \left(\frac{\alpha}{\pi} \right)^5 \left\{ 
\left(\frac{119}{243}-\frac{4}{81}\pi^2 \right)\log^3 \frac{m_{\mu}^2}{m_{e}^2} -
\left(\frac{61}{243}-\frac{2}{81}\pi^2 \right)\log^2 \frac{m_{\mu}^2}{m_{e}^2}
\right. \nn \\ & &  \hspace*{1.cm}
+\left( \frac{7627}{1458}+\frac{52}{81}\pi^2 -\frac{16}{135}\pi^4\right)\log \frac{m_{\mu}^2}{m_{e}^2} \nn \\ & &  \hspace*{1.cm}\left. +\frac{64244}{6561}-\frac{2593}{2187}\pi^2 +\frac{8}{405}\pi^4 -\frac{476}{81}\zeta(3)+\frac{16}{27}\pi^2\zeta(3)\right\}\,.
\eea}

\noi
This expression agrees with the one given by Laporta~\cite{La94}. We have also calculated the corresponding terms up to $\cO\left[
 \left(\frac{m_{e}^2}{m_{\mu}^2}\right)^{5/2}\right]$, as required by the wanted accuracy. Again,although numerically our result is much more precise than the one given by Kinoshita and Nio~\cite{KN04}, it agrees very well with it within their quoted errors.

	\item {\sc Two Electron Loops and Two Muon Loops}, Fig.~2(C) [six diagrams]:

	We recall the expression in Eq.~\rf{eemumu}	
	\be
	a_{\mu}^{(ee\mu\mu)}  =  \left(\frac{\alpha}{\pi}\right)^4 (-{\bf 6})\ \frac{1}{2\pi i}\int\limits_{c_s-i\infty}^{c_s+i\infty}ds \left(\frac{4m_e^2}{m_{\mu}^2} \right)^{-s}\Gamma(s)\Gamma(1-s)\ \Omega_2 (s)\ R_2 (s)\,,
\ee
with $\Omega_2 (s)$ and $R_2 (s)$ in Eqs.~\rf{omega2} and \rf{sigma2}. Here, the multipoles at $s=0$ generate the terms

{\setl
\bea
a_{\mu}^{(ee\mu\mu)} & \underset{{[s\ \ra\  0]}}{\thicksim}  & \left(\frac{\alpha}{\pi} \right)^5 \left\{\left(-\frac{943}{486}-\frac{8}{405}\pi^2 +\frac{16}{9}\zeta(3) \right)\log^2 \frac{m_{\mu}^2}{m_{e}^2}\right. \nn \\ & & \hspace*{1.2cm}  +\left(\frac{57899}{7290}-\frac{10766}{6075}\pi^2 +\frac{8}{81}\pi^4 -\frac{8}{135}\zeta(3) \right)\log\frac{m_{\mu}^2}{m_{e}^2}\nn \\ & & \hspace*{1.2cm}
\left. -\frac{1090561}{109350}-\frac{148921}{91125}\pi^2-\frac{106}{6075}\pi^4 +\frac{10732}{2025}\zeta(3)+\frac{32}{27}\pi^2 \zeta(3)\right\}\,,
\eea}

\noi
which agree with the expression found in Laporta~\cite{La94}. We have also calculated the contributions up to $\cO\left[\left(\frac{m_{e}^2}{m_{\mu}^2}\right)^3 \log^3 \frac{m_{\mu}^2}{m_{e}^2}\right]$, which can be found in the Appendix. Our numerical result agrees with the one given by Kinoshita and Nio~\cite{KN04} within their quoted errors.

	\item {\sc One Electron Loop and Three Muon Loops}, Fig 2.(D) [four diagrams]:
	
		We recall the expression in Eq.~\rf{emumumu}	
	
\be
	a_{\mu}^{(e\mu\mu\mu)}  = \left(\frac{\alpha}{\pi}\right)^5 (-{\bf 4})\ \frac{1}{2\pi i}\int\limits_{c_s-i\infty}^{c_s+i\infty}ds \left(\frac{4m_e^2}{m_{\mu}^2} \right)^{-s}\Gamma(s)\Gamma(1-s)\ \Omega_3 (s)\ R_1 (s) \,,
\ee
with $\Omega_3 (s)$ and $R_1 (s)$ in Eqs.~\rf{omega2} and \rf{sigma2}. Here, there is a double pole  at $s=0$ which generates the terms

{\setl
\bea
a_{\mu}^{(e\mu\mu\mu)} & \underset{{[s\ \ra\  0]}}{\thicksim}  & \left(\frac{\alpha}{\pi} \right)^5\left\{ \left(\frac{151849}{30618}-\frac{8}{135}\pi^4 +\frac{128}{189}\zeta(3)  \right)\log\frac{m_{\mu}^2}{m_{e}^2}  \right. \nn \\ & & \hspace*{1.2cm}
\left. -\frac{46796257}{3214890}+\frac{143}{81}\pi^2 +\frac{124}{8505}\pi^4 +\frac{92476}{6615}\zeta(3)-\frac{16}{9}\pi^2 \zeta(3)\right\}\,,
\eea}

\noi
in agreement wit the expression found in Laporta~\cite{La94}. We have also calculated the contributions up to $\cO\left[\left(\frac{m_{e}^2}{m_{\mu}^2}\right)^3 \log^3 \frac{m_{\mu}^2}{m_{e}^2}\right]$, which are given in the Appendix. Our numerical result also agrees very well with the one given by Kinoshita and Nio~\cite{KN04} within their quoted errors.

\end{itemize}
%%%%%%%%%%%%%%%%%%%%%%%%%%%%%%%%%%%%%%%%%%%%%%%%%%%%%%%%%%%%%%%%%%%%%%%%
%%%%%%%%%%%%%%%%%%%%%%%%%%%%%%%%%%%%%%%%%%%%%%%%%%%%%%%%%%%%%%%%%%%%%%%%
%%%%%%%%%%%%%%%%%%%%%%%%%%%%%%%%%%%%%%%%%%%%%%%%%%%%%%%%%%%%%%%%%%%%%%%%
%%%%%%%%%%%%%%%%%%%%%%%%%%%%%%%%%%%%%%%%%%%%%%%%%%%%%%%%%%%%%%%%%%%%%%%%
%%%%%%%%%%%%%%%%%%%%%%%%%%%%%%%%%%%%%%%%%%%%%%%%%%%%%%%%%%%%%%%%%%%%%%%%
%%%%%%%%%%%%%%%%%%%%%%%%%%%%%%%%%%%%%%%%%%%%%%%%%%%%%%%%%%%%%%%%%%%%%%%%
%%%%%%%%%%%%%%%%%%%%%%%%%%%%%%%%%%%%%%%%%%%%%%%%%%%%%%%%%%%%%%%%%%%%%%%%
%%%%%%%%%%%%%%%%%%%%%%%%%%%%%%%%%%%%%%%%%%%%%%%%%%%%%%%%%%%%%%%%%%%%%%%%

\section{\small Contributions from  Electron Loops and One Tau loop}
\setcounter{equation}{0}
\def\theequation{\arabic{section}.\arabic{equation}}

\noi
We shall finally discuss the calculation of vacuum polarization contributions involving electron loops and a tau loop. The corresponding expressions are the ones in Eqs.~\rf{eetau} and~\rf{eeetau}. As already mentioned, the problem here is the non factorization of the dependence in the the two Mellin variables $s$ and $t$ because of the presence in the integrand of the function $\Theta(s,t)$ defined in Eq.~\rf{theta}.

\subsection
 {\small Two Electron Loops and One Tau Loop} 
	
The corresponding expression in this case is the one given in Eq.~\rf{eetau}. Using the result obtained for  $R_{2}(s)$ in Eq.~\rf{sigma2} and the expression for $\Theta(s,t)$ in Eq.~\rf{theta}, we have:

{\setl
\bea\lbl{eetau2}
a_{\mu}^{(ee\tau)} & = & \left(\frac{\alpha}{\pi}\right)^4\frac{\bf 3}{(2i\pi)^2} \!\! \int\limits_{c_s-i\infty}^{c_s+i\infty} ds \int\limits_{c_s-i\infty}^{c_s+i\infty} dt \left(\frac{4m_e^2}{m_{\mu}^2} \right)^{-s} \left(\frac{m_\mu^2}{m_\tau^2} \right)^{-t}\frac{\Gamma(1+2s-2t)\ \Gamma(2-s+t)}{\Gamma(3+s-t)}\times \nn \\
& &  \frac{2}{9}\sqrt{\pi}\ \frac{6+13s+4s^2}{s^3 (2+s)(3+s)}\ \frac{\Gamma^2(1+s)\ \Gamma(2-s)}{\Gamma(\frac{3}{2}+s)}\ \   
 \frac{\Gamma(t)\ \Gamma(1-t)\ \Gamma^{2}(2-t)}{t\  \Gamma(4-2t)}\,.
\eea}

\noi
The evaluation of the asymptotic behaviour of this type of integral calls for a more sophisticated material than the {\it inverse mapping theorem} applied to the previous cases where there was only one ratio of masses.  Indeed, the generalization of the  {\it inverse mapping theorem} to this case is a typical problem of calculus of residues in $\mathbb{C}^2$. Let us then adopt the standard notation of multidimensional complex analysis~\footnote{See e.g., reference~\cite{GH78} for a comprehensive textbook.} and denote by $\omega^{(ee\tau)}$ the following $2$-form:

{\setl
\bea\lbl{omegamu}
\omega^{(ee\tau)} & = & \frac{2\sqrt{\pi}}{3}\left(\frac{4m_e^2}{m_{\mu}^2} \right)^{\!-s} \left(\frac{m_\mu^2}{m_\tau^2} \right)^{\!-t}     
 \frac{\Gamma(t)\ \Gamma(1-t)\ \Gamma^{2}(2-t)}{t\  \Gamma(4-2t)}\times \nn \\
& & \hspace*{4cm} \frac{(6+13s+4s^2)}{s^3 (2+s)(3+s)}\  \frac{\Gamma^2(s+1)\ \Gamma(2-s)}{\Gamma(s+\frac{3}{2})}  \times \nonumber \\
 & & \hspace*{4cm} \frac{\Gamma(1+2s-2t)\ \Gamma(2-s+t)}{\Gamma(3+s-t)}\  ds\wedge dt\,.
\eea}

\noi
Recall that the Mellin--Barnes representation in Eq.~\rf{eetau2} is valid for all $s$ and $t$ such that $\Ree (s)\in ]0,1[$ and $\Ree (t)\in ]-1,0[$; these two conditions resulting in the grey {\it fundamental square} in the plane $[ \Ree (s), \Ree (t) ]$, as illustrated in  Fig.~9. In the case of two Mellin variables, this  {\it fundamental square},  which in general may become a {\it fundamental polyhedra}, generalizes the concept of the {\it fundamental strip} in the case of one variable.
Since both $\frac{4m_e^2}{m_{\mu}^2} $ and $\frac{m_\mu^2}{m_\tau^2}$ are small, the asymptotic behaviour we are looking for in Eq.~\rf{eetau2}, will be  governed by the residues associated to the singularities of $\omega^{(ee\tau)}$ in the {\it cone} $\{\Ree (s)\leq 0, \Ree (t)\leq -1\}$, denoted by $\Pi$ in Fig.~9.
Formally, the solution to our problem is then given by the sum~\footnote{See ref.~\cite{AGdeR08} for technical details. Discussions in the literature of similar situations can be found e.g. in  Refs.~\cite{ZT98,PTZ94,PTC96}.} 

\be\lbl{formal}
a_{\mu}^{(ee\tau)}  =  \left(\frac{\alpha}{\pi}\right)^4 \sum_{(s_0,t_0)\in\Pi} \mathrm{Res}_{(s_0,t_0)}\, \omega^{(ee\tau)}\,.
\ee

From the expression in Eq.~\rf{omegamu}, one can easily see that the singularities in the plane  $[\Ree (s), \Ree (t) ]$  associated to $\omega^{(ee\tau)}$, are defined by a set of straight lines. For example, $\Gamma(2-s+t)$ induces a family of singular lines  parameterized by the affine equation: $\Ree (t)= \Ree (s) -(2+n),\,n\geq 0$. Each of the singular lines is called a {\it divisor}. As discussed in ref.~\cite{PTC96}, it is sufficient for our purposes to consider the {\it singular set of points} defined by the intersections of all the {\it divisors} in the appropriate {\it cone}. In our case, a subset of the nearest {\it singular points} to the origin of the {\it cone} $\Pi$, is illustrated by the red dots in  Fig.~9.

\begin{figure}[h]

\begin{center}
\includegraphics[width=0.3\textwidth]{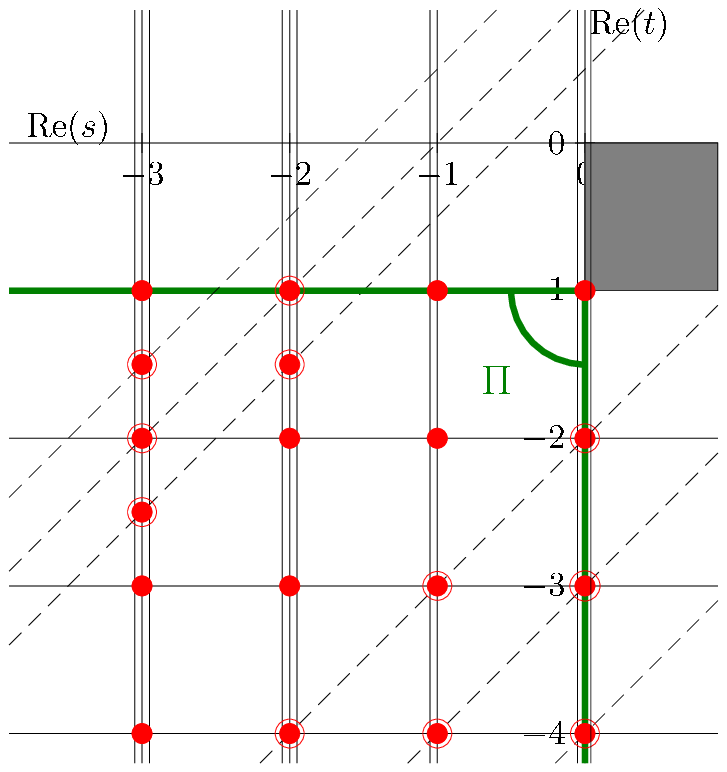}

\end{center}

\vspace*{0.25cm}
{\bf Fig.~9}
{\it\small  Plot of the divisors and the singular set of points of $\omega^{(ee\tau)}$ in Eq.~\rf{omegamu}. The small red dots  are singularities of the first kind; the larger red dots are singularities of the second kind. Notice that the singularities on the line $s=-\frac{2k+1}{2}$, $k\geq 1$, are screened by the presence of the factor $\Gamma(s+\frac{3}{2})$ in the denominator of $\omega^{(ee\tau)}$.}

\end{figure}

\noi
There are in fact two kinds of singularities to consider:

\begin{enumerate}

\item {\sc Singularities of the first kind,} {\it where only vertical and horizontal divisors (two or more)  cross each other.} 

In order to obtain the residue associated to such a singular point $(s_0,t_0)$, one proceeds as follows: 

\begin{itemize}

	\item  First perform the change of variables $s\leftrightarrow s_0+s$, $t\leftrightarrow t_0+t$ so as to bring the singularity to the origin $(0,0)$.
	
	\item  Use the functional relation $\Gamma(z+1)=z\Gamma(z)$ for each singular Gamma function until reaching an expression of the type
\be
\omega^{(ee\tau)}=\frac{h(s,t)}{s^nt^m}\ ds\wedge dt\,,
\ee
explicitly showing the full singular behaviour at the origin, 
with $h(s,t)$ an analytic function at $(0,0)$ and $n$ and $m$ positive integers.

 \item Then, use the Cauchy formula 
\be\lbl{firstkind}
\mathrm{Res}_{(0,0)}\, \omega^{(ee\tau)}=\frac{1}{(n-1)!(m-1)!}\frac{\partial^{n+m-2}\,h(s,t)}{\partial s^{n-1}\partial t^{m-1}}\Bigg\vert_{(0,0)}\,.
\ee
	
\end{itemize}

\item {\sc Singularities of the second kind,}  {\it where two or more divisors cross each other, one or more of them being oblique lines.}

This is the case, for instance, at the point  $(0,-2)$ in Fig.~9. In fact, in our case, the singularities of the second kind at a point $(s_0,t_0)$ are induced  either by the {\it oblique divisors} generated by
$\Gamma(1+2s +2t)$ or the {\it oblique divisors} generated by
$\Gamma(2-s +t)$.  
In order to obtain the residue associated to a singularity of the second kind at a generic point  $(s_0,t_0)$, one proceeds as follows: 

\begin{itemize}

\item 
Again, first perform the change of variables $s\leftrightarrow s_0+s$, $t\leftrightarrow t_0+t$ so as to bring the problem back to a singularity at the origin $(0,0)$; and then apply the functional relation $\Gamma(z+1)=z\Gamma(z)$ to the singular Gamma functions in question, until able to write $\omega^{(ee\tau)}$ in a way which explicitly shows the full singular behaviour in question:
\be\lbl{form}
\omega^{(ee\tau)}=\frac{h(s,t)}{s^n\,t^m\,(-s+t)}\ ds\wedge dt\,,
\ee
with $h(s,t)$ an analytic function at $(0,0)$ and $n$ and $m$ positive integers. The factor $t-s$ in the denominator is specific to the class of {\it oblique divisors} in our case which are all parallel lines to $s=t$ with multiplicity one.

\item
Apply the {\it Transformation Law for Residues}~\cite{GH78}, to the form $\omega^{(ee\tau)}$ in Eq.~\rf{form}  so as to fully decouple the $s$ and $t$ dependence in the denominator. This, of course, requires some explanation which we next provide. 

\end{itemize}

To be precise, let us discuss the case of the singular point $(0,-2)$ in Fig.~9.
There is a very important theorem in multidimensional complex analysis,  known as the {\it Transformation Law for Residues}~\cite{GH78} which, in the case of $\mathbb{C}^2$, states that if $U$ is an open set containing $(0,0)$; if $\mathbf{f}=\begin{bmatrix}f_1(s,t)\\f_2(s,t)\end{bmatrix}$ and $\mathbf{g}=\begin{bmatrix}g_1(s,t)\\g_2(s,t)\end{bmatrix}$ are two analytic mappings from $U$ to $\mathbb{C}^2$ such that $\mathbf{f}^{-1}(0,0)=\mathbf{g}^{-1}(0,0)=(0,0)$; and if there exists an analytic two-by-two matrix $A$ such that $\mathbf{g}=A \mathbf{f}$, then
\be\lbl{transfo}
\mathrm{Res}_{(0,0)}\,\frac{h(s,t)}{f_1(s,t)\ f_2(s,t)}\ ds \wedge dt=\mathrm{Res}_{(0,0)}\,\frac{h(s,t)\ \det A (s,t)}{g_1(s,t)\ g_2(s,t)}\ ds \wedge dt\,.
\ee
The application of this theorem to our case goes as follows:

\begin{itemize}
	\item 
First note that with the change of variables $s\ra s-t$, $t\ra t$, 
\be\lbl{modif}
\mathrm{Res}_{(0,0)}\,\frac{h(s,t)}{s^3t(-s+t)}\ ds\wedge dt=\mathrm{Res}_{(0,0)}\,\frac{h(-s+t,t)}{st(-s+t)^3}\ ds\wedge dt\,.
\ee
We do this because in this form and as shown in the next step, we can then  find easily a matrix $A$ to apply the {\it Transformation Law for Residues}. 

\item
Indeed, with $\mathbf{f}$ the vector: 
\be
\mathbf{f}=\begin{bmatrix}st\\(-s+t)^3\end{bmatrix}\,,
\ee
we can find a matrix $A$ which does the transformation:
\be
\underbrace{\begin{bmatrix}-s^4\\t^4\end{bmatrix}}_{\doteq\mathbf{g}}=\underbrace{\begin{bmatrix}-t^2+3st-3s^2&s\\s^2-3st+3t^2&t\end{bmatrix}}_{\doteq A} \underbrace{\begin{bmatrix}st\\ (-s+t)^3\end{bmatrix}}_{\doteq\mathbf{f}}\,,
\ee
and
\be
\det A=-(s^3+t^3)\,.
\ee

\item
By virtue of the theorem quoted above, we can then assert that

{\setl
\bea\lbl{intermed}
\mathrm{Res}_{(0,0)}\,\frac{h(-s+t,t)}{st(-s+t)^3}\ ds\wedge dt 
 & = & \mathrm{Res}_{(0,0)}\, h(-s+t,t)\frac{s^3+t^3}{s^4 t^4}\ ds\wedge dt \nn \\
 &=& \mathrm{Res}_{(0,0)}\, h(-s+t,t)\left(\frac{1}{st^4}+\frac{1}{s^4t}\right)\  ds\wedge dt \nonumber 
 \\
 &=& \frac{1}{6}\left[\frac{\partial^3 [h(-s+t,t)]}{\partial t^3}+\frac{\partial^3 [h(-s+t,t)]}{\partial s^3}\right]_{(0,0)}\,,
\eea}

\noi
where in going from the second line to the third, we have applied the Cauchy formula in Eq.~\rf{firstkind}.

\item
Finally, from Eq.~\rf{modif}, and applying the formula for chain--derivation  in Eq.~\rf{intermed}, we get the result
\be\lbl{secondkind}
\mathrm{Res}_{(0,0)}\,\frac{h(s,t)}{s^3t(-s+t)}\  ds\wedge dt = \frac{1}{2}\frac{\partial^3 h(s,t)}{\partial s^2\partial t}\Bigg\vert_{(0,0)}+\frac{1}{2}\frac{\partial^3 h(s,t)}{\partial s\partial t^2}\Bigg\vert_{(0,0)}+\frac{1}{6}\frac{\partial^3 h(s,t)}{\partial t^3}\Bigg\vert_{(0,0)} \,.
\ee

\end{itemize}
\end{enumerate}

We shall now proceed to the calculation of $a_{\mu}^{(ee\tau)}$ in Eq.~\rf{formal}, following the procedure which we have just outlined.

\begin{itemize}

\item

{\sc Singularities on the line $t=-1$}.

\begin{itemize}

\item The leading singularity is at the point 
 $(s_0,t_0)=(0,-1)$, and it is a  singularity  of the first kind. After translation to the origin and the reduction to an explicit singular form, we can write $\omega^{(ee\tau)}$ in the following way:
\be
\omega^{(ee\tau)}=\frac{h_{(0,-1)}(s,t)}{s^3t}\,ds\wedge dt\,,
\ee   
with

{\setl
\bea
h_{(0,-1)}(s,t) & = & \frac{2\sqrt{\pi}}{3}
\left(\frac{4m_e^2}{m_{\mu}^2} \right)^{-s} \left(\frac{m_\mu^2}{m_\tau^2} \right)^{1-t} \  \frac{\Gamma(3+2s-2t)\ \Gamma(1-s+t)}{\Gamma(4+s-t)}\times\nonumber \\
 & &  \frac{\Gamma^2(1+s) \Gamma(2-s)(6+13s+4s^2)}{(2+s)(s+3)\Gamma\left(\frac{3}{2}+s\right)}\ \   \frac{\Gamma^2(3-t) \Gamma(1+t) \Gamma(2-t)}{(-1+t)^2\ \Gamma(6-2t)}\,.
\eea}
 
 \noi
Then, using Eq.~\ref{firstkind}, we get

{\setl
\bea\lbl{leadingeetau}
\mathrm{Res}_{(0,-1)}\, \omega^{(ee\tau)} &=& \frac{1}{2}\frac{\partial^2 h_{(0,-1)}}{\partial s^2}\Bigg\vert_{(0,0)}\nonumber\\
 &=& \left(\frac{m_\mu^2}{m_\tau^2} \right)\left[\frac{1}{135}\log^2 \frac{m_\mu^2}{m_{e}^2} -\frac{1}{135}\log\frac{m_\mu^2}{m_{e}^2} -\frac{61}{2430}+\frac{2}{405}\pi^2\right]\,.
\eea}

\item The next singularity on the line $t=-1$ is at the point  $(s_0,t_0)=(-1,-1)$, and it is also a singularity of the first kind. In a similar way to the singularity in $(0,-1)$, one can easily find that the residue associated to  this singularity is given by the expression
\be\lbl{m1m1eetau}
\mathrm{Res}_{(-1,-1)}\, \omega^{(ee\tau)}=\left(\frac{m_e^2}{m_\tau^2}\right)\left[\frac{2}{15}\log\frac{ m_\mu^2}{m_e^2}-\frac{5}{9}\right]\,,
\ee
which gives a small contribution because of the suppression factor $\left(\frac{m_e^2}{m_\tau^2}\right)$.

\end{itemize}

\item

{\sc Singularities on the line $t=-2$}

\begin{itemize}

\item The singularity nearest to the origin on the line $t=-2$ is at $(s_0,t_0)=(0,-2)$ and it  is of the second kind. After translation to the origin and the reduction to an explicit singular form, we can write $\omega^{(ee\tau)}$ in the following way:
\be
\omega^{(ee\tau)}=\frac{h_{(0,-2)}(s,t)}{s^3t(-s+t)}\,ds\wedge dt\,,
\ee   
where

{\setl
\bea
h_{(0,-2)}(s,t) & = &
\left(\frac{4m_e^2}{m_{\mu}^2} \right)^{-s} \left(\frac{m_\mu^2}{m_\tau^2} \right)^{2-t} \frac{2\sqrt{\pi}}{3}\ 
\frac{\Gamma(5+2s-2t)\Gamma(1-s+t)}{\Gamma(5+s-t)}\times \nn \\
 & & \frac{(6+13s+4s^2)\Gamma^2(1+s)\Gamma(2-s)}{(2+s)(3+s)\Gamma\left(\frac{3}{2}+s\right)}\ 
\frac{\Gamma(1+t)\Gamma(3-t)}{(-1+t)(-2+t)^2}  \frac{\Gamma^2(4-t)}{\Gamma(8-2t)}\,.
\eea}

Then, using the result in Eq.~\ref{secondkind},we get

{\setl
\bea\lbl{0m2eetau}
\mathrm{Res}_{(0,-2)}\, \omega^{(ee\tau)} &=& \frac{1}{2}\frac{\partial^3 h_{(0,-2)}}{\partial s^2\partial t}|_{(0,0)}+\frac{1}{2}\frac{\partial^3 h_{(0,-2)}}{\partial s\partial t^2}|_{(0,0)}+\frac{1}{6}\frac{\partial^3 h_{(0,-2)}}{\partial t^3}|_{(0,0)} \nonumber \\
 & & \nn \\
 &=& \left(\frac{m_\mu^2}{m_\tau^2} \right)^2 \left[\frac{1}{1260}\log^3\frac{m_\mu^2}{m_\tau^2}   -\left(\frac{1}{420}\log\frac{m_\mu^2}{m_e^2}+\frac{37}{44100}\right)\log^2\frac{m_\mu^2}{m_\tau^2} \right. \nonumber \\
 & & +\left(\frac{1}{420}\log^2\frac{m_\mu^2}{m_e^2}   
+\frac{37}{22050}\log\frac{m_\mu^2}{m_e^2}+\frac{40783}{4630500}\right) \log\frac{m_\mu^2}{m_\tau^2}   \nonumber \\
 & & +\frac{3}{19600}\log^2\frac{m_\mu^2}{m_e^2}  +\left(\frac{\pi^2}{630}-\frac{229213}{12348000}\right)\log\frac{m_\mu^2}{m_e^2}
 \nn \\
  & & \left.  +\frac{\pi^2}{1512}-\frac{30026659}{5186160000}\right]\,,
\eea}

\noi 
which for convenience we have ordered in decreasing powers of  $\left(\frac{m_\mu^2}{m_\tau^2} \right)^2 \log^n\frac{m_\mu^2}{m_\tau^2}$ with $n=3,2,1,0$.

\end{itemize}

The other singularities on the line $t=-2$ (for $s=-1,\,-2$ and further) turn out to give contributions which are too small  to be of physical relevance.

\item

{\sc Other Singularities}

In fact there is only one more singularity which can give rise to a term of the order of the present experimental error limitations. This is the singularity	at $(0,-3)$. Its residue can be computed exactly in the same way as the one for the singularity at $(0,-2)$ with the result, ordered in decreasing powers of  $\left(\frac{m_\mu^2}{m_\tau^2} \right)^3 \log^n\frac{m_\mu^2}{m_\tau^2}$ with $n=3,2,1,0$:

{\setl
\bea\lbl{0m3eetau}
\mathrm{Res}_{(0,-3)}\, \omega^{(ee\tau)} &=& \left(\frac{m_\mu^2}{m_\tau^2}\right)^3  \left[\frac{2}{2835}\log^3\frac{m_\mu^2}{m_\tau^2}  - \left(\frac{2}{945}\log\frac{m_\mu^2}{m_e^2}+\frac{199}{595350}\right)\log^2\frac{m_\mu^2}{m_\tau^2} \right. \nonumber \\
 & & +\left(\frac{2}{945}\log^2\frac{m_\mu^2}{m_e^2}   +\frac{199}{297675}\log\frac{m_\mu^2}{m_e^2} 
+\frac{1368473}{187535250}\right)\log\frac{m_\mu^2}{m_\tau^2} \nonumber \\
 & & + \frac{131}{297675}\log^2\frac{m_\mu^2}{m_e^2}  -\left(\frac{4}{2835}\pi^2-\frac{1102961}{75014100}\right)\log\frac{m_\mu^2}{m_e^2} \nn \\ 
  & & \left. +\frac{\pi^2}{14175}-\frac{311791591}{472588830000}\right]\,.
\eea} 

\noi
The total contribution from Eqs.~\rf{leadingeetau}, \rf{m1m1eetau}, \rf{0m2eetau} and \rf{0m3eetau} to $a_{\mu}^{(ee\tau)}$, ordered in a decreasing order of magnitude,  is given in Eq.~\rf{eetau} in the Appendix.

\end{itemize}

\subsection
 {\small Three Electron Loops and One Tau Loop}	
 
 The corresponding expression in this case is the one given in Eq.~\rf{eeetau}. Using the result obtained for  $R_{3}(s)$ in Eq.~\rf{sigma3} and the expression for $\Theta(s,t)$ in Eq.~\rf{theta}, we can write explicitly the two form  associated to this integral as follows:

{\setl
\bea\lbl{omegamueeetau}
\omega^{(eee\tau)} &=& \frac{8\sqrt{\pi}}{864}\left(\frac{4m_e^2}{m_{\mu}^2} \right)^{-s} \left(\frac{m_\mu^2}{m_\tau^2} \right)^{-t}\  \frac{\Gamma(t)\ \Gamma(1-t)\ \Gamma^{2}(2-t)}{t\  \Gamma(4-2t)}\times \nn \\ 
 & &   
 \frac{\Gamma^2(s)\Gamma(1-s)}{\Gamma\left(\frac{11}{2}+s\right)}
\left[\frac{P_7(s)}{s(1+s)(2+s)}-(1+s)(35+21s+3s^2)\left(27\pi^2-162\  \psi^{(1)}(s)\right) \right]\times\nonumber \\
 & & \hspace*{4.5cm} \    
\frac{\Gamma(1+2s-2t)\ \Gamma(2-s+t)}{\Gamma(3+s-t)}\  ds\wedge dt\,,
\eea}

\noi
where $P_7(s)$ is the polynomial given in Eq.~\rf{pol7}.
When comparing the two forms $\omega_\mu^{(ee\tau)}$ in Eq.~\rf{omegamu} and $\omega_\mu^{(eee\tau)}$ in Eq.~\rf{omegamueeetau}, we can see that they only differ in the form of the factorized $s$--dependence. Therefore, except for the multiplicity of the vertical {\it divisors}, the plot of the {\it singular points} associated to $\omega_\mu^{(eee\tau)}$ which we show  in Fig.~10 is pretty much the same as the one for $\omega_\mu^{(eee\tau)}$ in Fig.~9. It is then not surprising that the calculation of the asymptotic behaviour of the integral in  Eq.~\rf{eeetau} turns out to be very similar to the one discussed in the previous subsection.    

\begin{figure}[h]

\begin{center}
\includegraphics[width=0.3\textwidth]{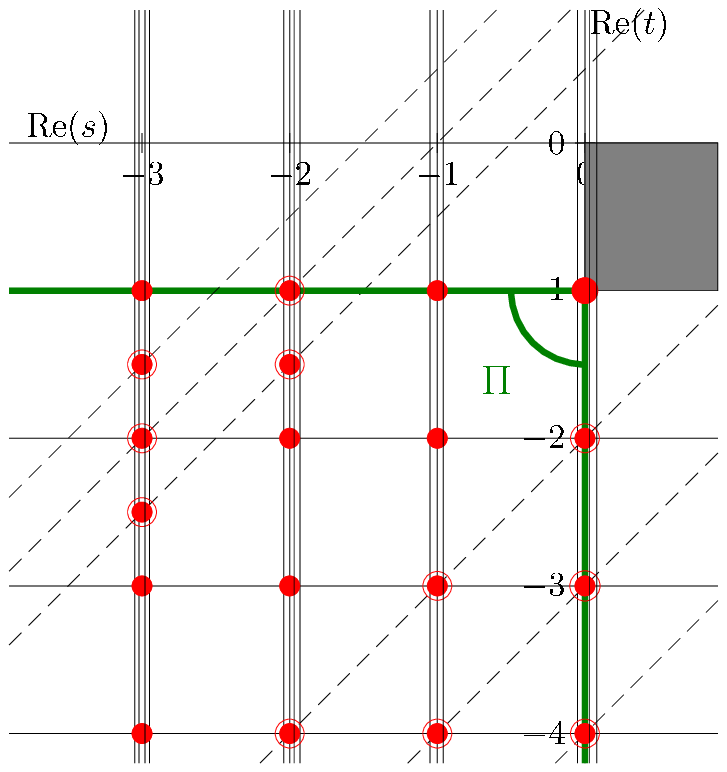}

\end{center}

\vspace*{0.25cm}
{\bf Fig.~10}
{\it\small  Plot of some of the divisors and singular points of $\omega_\mu^{(eee\tau)}$ with the same conventions as in Fig.~9. Only the multiplicity of the vertical divisors differs from the former case.
}

\end{figure}

Residues at the {\it singular points} in Fig.~10 are  computed following the strategy discussed in the previous subsection, where one uses the functional relations: $\Gamma(z+1)=z\Gamma(z)$ and $\psi^{(1)}(1+z)=\psi^{(1)}(z)-\frac{1}{z^2}$.

\noi
For the singularities of the first kind one can then use Eq.~\rf{firstkind}. The new type of singularities of the second kind are reducible
to a form of the type
\be\lbl{form2}
\omega^{(eee\tau)}=\frac{\mathsf{h}(s,t)}{s^4\,t\,(-s+t)}\,ds\wedge dt\,,
\ee
with $\mathsf{h}$ 
some analytic function at the origin. In this case, the appropriate $A$-matrix to implement the {\it Transformation Law for Residues}   reads:
\be
\begin{bmatrix}
-t^3+4st^2-6s^2t+4s^3 & s \\
-s^3+4s^2t-6st^2+4t^3 & t
\end{bmatrix}\,,
\ee
and proceeding as explained in the previous subsection one finally obtains:
\be\lbl{secondkind2}
\mathrm{Res}_{(0,0)} \frac{\mathsf{h}(s,t)}{s^4\,t\,(-s+t)}\,ds\wedge dt = \frac{1}{4!}\left[\frac{\partial^4 \mathsf{h}(s,t)}{\partial t^4}+4\frac{\partial^4 \mathsf{h}(s,t)}{\partial s\partial t^3}+6\frac{\partial^4 \mathsf{h}(s,t)}{\partial s^2 \partial t^2}+4\frac{\partial^3 \mathsf{h}s,t)}{\partial s^3\partial t}\right]_{(0,0)}\,.
\ee
 
Let us now discuss the explicit calculation of $a_{\mu}^{(eee\tau)}$. 

\begin{itemize}

\item

{\sc Singularities on the line $t=-1$}.

\begin{itemize}

\item As in the case of $a_{\mu}^{(ee\tau)}$, the leading singularity is the one at  $(s_0,t_0)=(0,-1)$ and it is a singularity of the first kind. Performing the change of variables $t\leftrightarrow -1+t$ we write:
\be
\omega^{(eee\tau)}=\frac{\mathsf{h}_{(0,-1)}(s,t)}{s^4t}\,ds\wedge dt\,,
\ee 
where

{\setl
\bea\lbl{h0m1}
\mathsf{h}_{(0,-1)}(s,t) &=& \frac{8\sqrt{\pi}}{864}\left(\frac{4m_e^2}{m_{\mu}^2} \right)^{-s} \left(\frac{m_\mu^2}{m_\tau^2} \right)^{1-t}\  \frac{\Gamma(1+t)\ \Gamma(2-t)\ \Gamma^{2}(3-t)}{(-1+t)^2\  \Gamma(6-2t)} \frac{\Gamma^2(1+s)\Gamma(1-s)}{\Gamma\left(\frac{11}{2}+s\right)}\times \nn \\ 
 & &   
\hspace*{-1cm}
\left\{\frac{s\ P_7(s)}{(1+s)(2+s)}-(1+s)(35+21s+3s^2)\left[-162+27\pi^2 s^2 -162\  \psi^{(1)}(1+s)\right] \right\}\times\nonumber \\
 & & \hspace*{4cm} \    
\frac{\Gamma(3+2s-2t)\ \Gamma(1-s+t)}{\Gamma(4+s-t)}\,.
\eea}

From this expression we obtain

{\setl
\bea\lbl{leadingeeetau}
\mathrm{Res}_{(0,-1)}\, \omega^{(eee\tau)} &=& \frac{1}{3!}\frac{\partial^3 \mathsf{h}_{(0,-1)}}{\partial s^3}\Bigg\vert_{(0,0)}\nonumber\\
 &=& \left(\frac{m_\mu^2}{m_\tau^2} \right)\left[\frac{4}{1215}\log^3\frac{m_\mu^2}{m_{e}^2} -\frac{2}{405}\log^2 \frac{m_\mu^2}{m_{e}^2} -\left(\frac{122}{3645}-\frac{8\pi^2}{1215}\right)\log\frac{m_\mu^2}{m_{e}^2} \right. \nonumber\\
 &+& \left. \frac{2269}{32805}-\frac{4\pi^2}{215}-\frac{16}{405}\zeta(3)\right]\,,
\eea}

\noi
which provides the leading contribution to $a_{\mu}^{(eee\tau)}$

\item
The next singularity is at $(s_0,t_0)=(-1,-1)$. Also a singularity of the first kind, whose residue we find to be
\be\lbl{subleadingeeetau}
\mathrm{Res}_{(-1,-1)}\, \omega^{(eee\tau)} =
\left(\frac{m_e^2}{m_\tau^2} \right)\left[\frac{4}{45}\log^2 \frac{m_\mu^2}{m_{e}^2} -\frac{20}{27}\log\frac{m_\mu^2}{m_{e}^2} +\frac{634}{405}+\frac{8\pi^2}{135}\right]\,,
\ee
and gives a contribution suppressed by a factor $\left(\frac{m_e^2}{m_\tau^2} \right)$.

\end{itemize}

\item  

{\sc Singularities on the line $t=-2$}.

\begin{itemize}

\item

The singularity at $(s_0,t_0)=(0,-2)$ is  of the second kind. After the change of variables $t\leftrightarrow -2+t$, and using standard functional relations, one can write $\omega^{(eee\tau)}$ in the following way 
\be
\omega^{(eee\tau)}=\frac{\mathsf{h}_{(0,-2)}(s,t)}{s^4t(-s+t)}\,ds\wedge dt
\ee
with

{\setl
\bea\lbl{h0m2}
\mathsf{h}_{(0,-2)}(s,t) &=& \frac{8\sqrt{\pi}}{864}\left(\frac{4m_e^2}{m_{\mu}^2} \right)^{-s} \left(\frac{m_\mu^2}{m_\tau^2} \right)^{2-t}\  \frac{\Gamma(1+t)\ \Gamma(3-t)\ \Gamma^{2}(4-t)}{(-1+t)(-2+t)^2\  \Gamma(8-2t)} \frac{\Gamma^2(1+s)\Gamma(1-s)}{\Gamma\left(\frac{11}{2}+s\right)}\times \nn \\ 
 & &   
\hspace*{-1cm}
\left\{\frac{s\ P_7(s)}{(1+s)(2+s)}-(1+s)(35+21s+3s^2)\left[-162+27\pi^2 s^2 -162\  \psi^{(1)}(1+s)\right] \right\}\times\nonumber \\ & & \hspace*{4cm} \    
\frac{\Gamma(5+2s-2t)\ \Gamma(1-s+t)}{\Gamma(5+s-t)}\,.
\eea}

\noi
According to Eq.~\rf{secondkind2}, we have:

{\setl
\bea\lbl{secondordereeetau}
\mathrm{Res}_{(0,-2)}\, \omega^{(eee\tau)} &=& \frac{1}{4!}\left[\frac{\partial^4 \mathsf{h}_{(0,-2)}}{\partial t^4}+4\frac{\partial^4 \mathsf{h}_{(0,-2)}}{\partial s\partial t^3}+6\frac{\partial^4 \mathsf{h}_{(0,-2)}}{\partial s^2\partial t^2}+4\frac{\partial^3 \mathsf{h}_{(0,-2)}}{\partial s^3\partial t}\right]_{(0,0)}\nonumber\\
 &=& \left(\frac{m_\mu^2}{m_\tau^2} \right)^2\left[ -\frac{1}{3780}\log^4\frac{m_\mu^2}{m_{\tau}^2} -\left(\frac{1}{945}\log\frac{m_\mu^2}{m_{e}^2}-\frac{37}{99225}\right)\log^3\frac{m_\mu^2}{m_{\tau}^2} \right. \nonumber\\
 &-& \left(\frac{1}{630}\log^2\frac{m_\mu^2}{m_e^2}+\frac{37}{33075}\log\frac{m_\mu^2}{m_e^2}-\frac{40783}{6945750}\right)\log^2\frac{m_\mu^2}{m_{\tau}^2}  \nonumber \\
 &+&  \left(\frac{1}{945}\log^3\frac{m_\mu^2}{m_e^2}+\frac{37}{33075}\log^2\frac{m_\mu^2}{m_e^2}-\frac{40783}{3472875}\log\frac{m_\mu^2}{m_e^2}-\frac{4}{315}\zeta(3)\right)\log\frac{m_\mu^2}{m_{\tau}^2}  \nonumber\\
 &+& \frac{1}{14700}\log^3\frac{m_\mu^2}{m_{e}^2}+\left(\frac{229213}{18522000}+\frac{\pi^2}{945}\right)\log^2\frac{m_\mu^2}{m_e^2}  \nonumber\\
 &-& \left(\frac{30026659}{3889620000}-\frac{\pi^2}{1134}\right)\log\frac{m_\mu^2}{m_e^2}  \nonumber\\
 &-& \left. \frac{24827672279}{544546800000} -\frac{59}{13608}\pi^2+\frac{4}{4725}\pi^4-\frac{1}{1225}\zeta(3)\right]
\eea}

\noi
which for convenience we have ordered in decreasing powers of  $\left(\frac{m_\mu^2}{m_\tau^2} \right)^2 \log^n\frac{m_\mu^2}{m_\tau^2}$ with $n=4,3,2,1,0$.
%%%%%%%%%%%%%%%
%%%%%%%%%%%%%%%

\end{itemize}

\item

{\sc Singularities on the line $t=-3$}.

\begin{itemize}

\item

The singularity at $(0,-3)$ is also of the second kind, and its residue can be computed in the same way as the one for the singularity at $(0,-2)$. One finds:

{\setl
\bea\lbl{thirdordereeetau}
\mathrm{Res}_{(0,-3)}\, \omega^{(eee\tau)}  &=& \left(\frac{m_\mu^2}{m_\tau^2} \right)^3\left[ -\frac{2}{8505}\log^4\frac{m_\mu^2}{m_{\tau}^2} -\left(\frac{8}{8505}\log\frac{m_\mu^2}{m_{e}^2}-\frac{398}{2679075}\right)\log^3\frac{m_\mu^2}{m_{\tau}^2} \right. \nonumber\\
 &-& \left(\frac{4}{2835}\log^2\frac{m_\mu^2}{m_e^2}+\frac{398}{893025}\log\frac{m_\mu^2}{m_e^2}-\frac{1368473}{281302875}\right)\log^2\frac{m_\mu^2}{m_{\tau}^2}  \nonumber \\
 &+&  \left(\frac{8}{8505}\log^3\frac{m_\mu^2}{m_e^2}+\frac{398}{893025}\log^2\frac{m_\mu^2}{m_e^2}-\frac{2736946}{281302875}\log\frac{m_\mu^2}{m_e^2}-\frac{32}{2835}\zeta(3)\right)\log\frac{m_\mu^2}{m_{\tau}^2}  \nonumber\\
 &-& \frac{524}{2679075}\log^3\frac{m_\mu^2}{m_{e}^2}+\left(\frac{1102961}{112521150}+\frac{8\pi^2}{8505}\right)\log^2\frac{m_\mu^2}{m_e^2}  \nonumber\\
 &-& \left(\frac{311791591}{254441622500}+\frac{4\pi^2}{42525}\right)\log\frac{m_\mu^2}{m_e^2}  \nonumber\\
 &-& \left. \frac{20302969165147}{446596444350000} -\frac{6299}{1913625}\pi^2+\frac{32}{42525}\pi^4+\frac{2096}{893025}\zeta(3)\right]
\eea}

\noi
which for convenience we have ordered in decreasing powers of  $\left(\frac{m_\mu^2}{m_\tau^2} \right)^3 \log^n\frac{m_\mu^2}{m_\tau^2}$ with $n=4,3,2,1,0$.

\end{itemize}

The other singularities have residues which give contributions smaller than the error induced by the leading term.
The total contribution from Eqs.~\rf{leadingeeetau}, \rf{subleadingeeetau}, \rf{secondordereeetau} and \rf{thirdordereeetau} to $a_{\mu}^{(eee\tau)}$, ordered in a decreasing order of magnitude, is given in Eq.~\rf{Aeeetau} in the Appendix. Notice that, numerically, the contribution with one tau loop and three electron loops:  $a_{\mu}^{(eee\tau)}$, is of the same size as the contribution from three muon loops and one electron loop: $a_{\mu}^{(e\mu\mu\mu)}$. 

\end{itemize}

%%%%%%%%%%%%%%%%
%%%%%%%%%%%%%%%%
%%%%%%%%%%%%%%%%

\begin{appendix}
\renewcommand{\thesubsection}{\normalsize \Alph {subsection}}

\begin{center}
{\bf\normalsize APPENDIX}
\end{center}

\noi The purpose of this appendix is to collect systematically the various results discussed in the previous sections. This may be useful to  readers who are only interested in the final analytic expressions and the numerical results. The values that we have used for the lepton masses are the ones in the 2006 PDG booklet~\cite{PDG06}:
\be
m_e=0.510~998~92(04)~\MeV\,,\quad m_{\mu}=105.658~369(9)~\MeV\quad\annd\quad
m_{\tau}=1776.99(29)~\MeV\,.
\ee

\subsection{\small Eighth Order Results}
\setcounter{equation}{0}
\def\theequation{\Alph{subsection}.\arabic{equation}}

 These are the results corresponding to the Feynman diagrams in Fig.~1 with the combinatoric factors included. The numbers in italics  are the results from  the numerical evaluation of Kinoshita and Nio in ref.~\cite{KN04}.
 
%\vspace*{0.5cm}
{\footnotesize
{\setl
\bea
a_{\mu}^{(eee)} & = & \left(\frac{\alpha}{\pi} \right)^4 \left\{\frac{1}{54} \log^3 \frac{m_{\mu}^2}{m_{e}^2}  -\frac{25}{108} \log^2 \frac{m_{\mu}^2}{m_{e}^2}  +\left(\frac{317}{324}+\frac{\pi^2}{27}\right) \log \frac{m_{\mu}^2}{m_{e}^2} 
  -\frac{8609}{5832}-\frac{25}{162}\pi^2-\frac{2}{9}\zeta(3)\right. \nn \\
 &  & +\left(\frac{m_{e}^2}{m_{\mu}^2}\right)^{1/2} \ \frac{101}{1536}\pi^4 \nn \\
 & &  + \left(\frac{m_{e}^2}{m_{\mu}^2}\right)\ \left[
-\frac{2}{9}\log^3 \frac{m_{\mu}^2}{m_{e}^2} +\frac{13}{9}\log^2 \frac{m_{\mu}^2}{m_{e}^2}  -\left( \frac{152}{27}+\frac{4}{9}\pi^2\right)\log \frac{m_{\mu}^2}{m_{e}^2}   +\frac{967}{315}+\frac{26}{27}\pi^2+\frac{136}{35}\zeta(3)\right] \nn \\
& &  +\left(\frac{m_{e}^2}{m_{\mu}^2}\right)^{3/2} \ \left(-\frac{205}{384}\pi^4\right) \nn \\
& & + \left(\frac{m_{e}^2}{m_{\mu}^2}\right)^2\ \left[\frac{1}{12}\log^4 \frac{m_{\mu}^2}{m_{e}^2}+\frac{8}{27}\log^3 \frac{m_{\mu}^2}{m_{e}^2} +\left(\frac{127}{108}+\frac{\pi^2}{3}\right)\log^2 \frac{m_{\mu}^2}{m_{e}^2} 
+\left(\frac{236}{27}+\frac{16}{27}\pi^2-4\zeta(3) \right)\log \frac{m_{\mu}^2}{m_{e}^2} \right. \nn \\
 & & \hspace*{1.8cm} \left.\left. +\frac{63233}{3240}+\frac{127}{162}\pi^2+\frac{\pi^4}{5}-\frac{64}{15}\zeta(3)\right]
+\cO
 \left(\frac{m_{e}^2}{m_{\mu}^2}\right)^{5/2} \right\}\\
  & & \nn \\
   & = & \left(\frac{\alpha}{\pi} \right)^4  \ 7.223~076~98(14)\quad  [{\it 7.223~077(29)}]\,. \lbl{Aeee}
\eea} } 

%\vspace*{0.5cm}
{\footnotesize
{\setl
\bea
a_{\mu}^{(ee\mu)} & = & \left(\frac{\alpha}{\pi} \right)^4 \left\{\left(\frac{119}{108}-\frac{\pi^2}{9}\right) \log^2 \frac{m_{\mu}^2}{m_{e}^2}-\left(\frac{61}{162}-\frac{\pi^2}{27} \right)\log \frac{m_{\mu}^2}{m_{e}^2}+\frac{7627}{1944}+\frac{13}{27}\pi^2-\frac{4}{45}\pi^4 \right. \nn \\
 & &  +\left(\frac{m_{e}^2} {m_{\mu}^2}\right) 
\left[ \left(-\frac{115}{27}+\frac{4}{9}\pi^2 \right) \log\frac{m_{\mu}^2}{m_{e}^2}+
\frac{227}{18}-\frac{4}{3}\pi^2\right]\nn \\
 & &  + \left(\frac{m_{e}^2} {m_{\mu}^2}\right)^2 \left[\frac{4}{45}\log^3\frac{m_{\mu}^2}{m_{e}^2} +\left(\frac{863}{450}-\frac{2}{9}\pi^2\right)\log^2\frac{m_{\mu}^2}{m_{e}^2} +
 \left( \frac{268061}{40500}-\frac{13}{27}\pi^2\right)\log\frac{m_{\mu}^2}{m_{e}^2}\right. \nn \\
 & & \hspace*{1.6cm}\left.\left. +\frac{9200857}{1215000}+\frac{67}{81}\pi^2-\frac{8}{45}\pi^4 \right]+
 \cO\left[
 \left(\frac{m_{e}^2}{m_{\mu}^2}\right)^{3}\log^2\frac{m_{\mu}^2}{m_{e}^2}\right] \right\} \\
 & & \nn \\
  & = & \left(\frac{\alpha}{\pi} \right)^4 \  0.494~072~046(5)\quad [\it 0.494~075(6)]\,.
\eea}
}

%\vspace*{0.5cm}
{\footnotesize
{\setl
\bea
a_{\mu}^{(e\mu\mu)} & = & \left(\frac{\alpha}{\pi}\right)^4 \left\{\left[\left(-\frac{943}{324}-\frac{4}{135}\pi^2+\frac{8}{3}\zeta(3) \right)\log \frac{m_{\mu}^2}{m_{e}^2}
 +\frac{57899}{9720}-\frac{5383}{4050}\pi^2+\frac{2}{27}\pi^4-\frac{2}{45}\zeta(3) \right]\right. \nn \\
  & &  +\left(\frac{m_{e}^2}{m_{\mu}^2}\right)\left[\frac{458}{81}-\frac{26}{105}\pi^2-\frac{8}{3}\zeta(3) \right]\nn \\
   & &   +\left(\frac{m_{e}^2}{m_{\mu}^2}\right)^2 \left[\left(-\frac{235}{486}-\frac{65}{567}\pi^2+\frac{4}{3}\zeta(3) \right)\log \frac{m_{\mu}^2}{m_{e}^2}\right.\nn \\
    & & \hspace*{1.6cm}\left.\left. -\frac{93851}{122472}-\frac{257449}{357210}\pi^2 +\frac{1}{27}\pi^4 +\frac{676}{189}\zeta(3) \right]   +\cO
 \left[\left(\frac{m_{e}^2}{m_{\mu}^2}\right)^{3}\log \frac{m_{\mu}^2}{m_{e}^2}\right]        \right\} \\
 & & \nn \\
  & = & \left(\frac{\alpha}{\pi} \right)^4 \ 0.027~988~322~7(1)\quad [\it 0.027~988(1)]\,.
\eea}}

%\vspace*{0.5cm}
{\footnotesize
{\setl
\bea
a_{\mu}^{(ee\tau)} & = & \left(\frac{\alpha}{\pi}\right)^4 \left\{
\left(\frac{m_\mu^2}{m_\tau^2} \right)\left[\frac{1}{135}\log^2 \frac{m_\mu^2}{m_{e}^2} -\frac{1}{135}\log\frac{m_\mu^2}{m_{e}^2} -\frac{61}{2430}+\frac{2}{405}\pi^2\right] \right. \nn \\
 & & \hspace*{0.8cm}+
\left(\frac{m_\mu^2}{m_\tau^2} \right)^2 \left[-\frac{1}{420}\log\frac{m_\tau^2}{m_e^2}\log\frac{m_\mu^2}{m_e^2}\log\frac{m_\tau^2}{m_\mu^2}-\frac{1}{1260}\log^3\frac{m_\tau^2}{m_\mu^2}\right. \nn \\
 & & \hspace*{1.5cm}
+\frac{3}{19600}\log^2\frac{m_\mu^2}{m_e^2} -\frac{37}{22050}\log\frac{m_\mu^2}{m_e^2}\log\frac{m_\tau^2}{m_\mu^2}
+\frac{37}{44100}\log^2\frac{m_\tau^2}{m_\mu^2}\nn \\
 & & \hspace*{1.5cm}
-\left(\frac{229213}{12348000}-\frac{\pi^2}{630}\right)\log\frac{m_\mu^2}{m_e^2}
-\frac{40783}{4630500} \log\frac{m_\tau^2}{m_\mu^2} \nn \\
 & & \hspace*{1.5cm} \left. -\frac{30026659}{5186160000}+\frac{\pi^2}{1512}\right]
 \nn \\
  & & \hspace*{0.8cm} +\left(\frac{m_e^2}{m_\tau^2}\right)\left[\frac{2}{15}\log\frac{ m_\mu^2}{m_e^2}-\frac{5}{9}\right]\nn \\
   & & 
 \hspace*{0.8cm} +  \left(\frac{m_\mu^2}{m_\tau^2}\right)^3 \left[-\frac{2}{945}\log\frac{m_\tau^2}{m_e^2}\log\frac{m_\mu^2}{m_e^2} \log\frac{m_\tau^2}{m_\mu^2}-\frac{2}{2835}\log^3\frac{m_\tau^2}{m_\mu^2}\right.
 \nn \\
 & &  \hspace*{1.5cm}  + \frac{131}{297675}\log^2\frac{m_\mu^2}{m_e^2}  
 -\frac{199}{297675}\log\frac{m_\mu^2}{m_e^2} \log\frac{m_\tau^2}{m_\mu^2} 
 -\frac{199}{595350}\log^2\frac{m_\tau^2}{m_\mu^2}  \nn \\
  & &  \hspace*{1.5cm}   -\left(\frac{1102961}{75014100}-\frac{4}{2835}\pi^2\right)\log\frac{m_\mu^2}{m_e^2}   -\frac{1368473}{187535250}\log\frac{m_\tau^2}
{m_\mu^2}               \nn \\
& & \hspace*{1.5cm} \left. -\frac{311791591}{472588830000}-\frac{\pi^2}{14175}\right] \nn  \\
 & & \left. +_ \cO\left[ \left(\frac{m_\mu^2}{m_\tau^2}\right)^4 \log\frac{m_\tau^2}{m_e^2}\log\frac{m_\mu^2}{m_e^2} \log\frac{m_\tau^2}{m_\mu^2}\right] \right\}\\
 & & \nn \\
  & = & \left(\frac{\alpha}{\pi} \right)^4 \ 0.002~748~6(9)\,.
\eea}}

\subsection{\small Tenth Order Results}
\setcounter{equation}{0}
\def\theequation{\Alph{subsection}.\arabic{equation}}

 These are the results corresponding to the Feynman diagrams in Fig.~2 with the combinatoric factors included. Our numerical results for $a_{\mu}^{(eeee)}$, $a_{\mu}^{(eee\mu)}$, $a_{\mu}^{(ee\mu\mu)}$, and $a_{\mu}^{(e\mu\mu\mu)}$   agree, within errors, with those of Laporta~\cite{La94} which, however, he obtained using an older determination of the mass ratio  $\frac{m_{\mu}}{m_{e}}=206.768~262(30)$. The number in italics are the results from  the numerical evaluation of Kinoshita and Nio in ref.~\cite{KN06}.
 
 %\vspace*{0.5cm}
{\footnotesize
{\setl
\bea
a_{\mu}^{(eeee)} & = & \left(\frac{\alpha}{\pi} \right)^5 \left\{\frac{1}{162}\log^4 \frac{m_{\mu}^2}{m_{e}^2}-\frac{25}{243}\log^3 \frac{m_{\mu}^2}{m_{e}^2}+
\left(\frac{317}{486}+\frac{2}{81}\pi^2 \right) \log^2 \frac{m_{\mu}^2}{m_{e}^2}
-\left(\frac{8609}{4374}+\frac{50}{243}\pi^2+\frac{8}{27}\zeta(3) \right)\log \frac{m_{\mu}^2}{m_{e}^2}\right. \nn \\
& & \hspace*{1.cm}  +\frac{64613}{26244}+\frac{317}{729}\pi^2+\frac{2}{135}\pi^4+\frac{100}{81}\zeta(3)\nn \\ 
 & &  + \left(\frac{m_{e}^2}{m_{\mu}^2}\right)^{1/2}\left(-\frac{18203}{374220}\pi^4 \right)\nn \\
  & & +\left(\frac{m_{e}^2}{m_{\mu}^2}\right)\left[\frac{2}{27}\log^4 \frac{m_{\mu}^2}{m_{e}^2}+\frac{52}{81}\log^3 \frac{m_{\mu}^2}{m_{e}^2}-\left(\frac{304}{81}+\frac{8}{27}\pi^2 \right)\log^2 \frac{m_{\mu}^2}{m_{e}^2}+\left( \frac{4924}{729}+\frac{104}{81}\pi^2 +\frac{32}{9}\zeta(3)\right)\log \frac{m_{\mu}^2}{m_{e}^2}\right. \nn \\ 
  & & \left.\hspace*{1.6cm} -\frac{55766}{6075}
  -\frac{608}{243}\pi^2 -\frac{8}{45}\pi^4 -\frac{592}{75}\zeta(3)\right] \nn \\
   & &   +\left(\frac{m_{e}^2}{m_{\mu}^2}\right)^{3/2}\left(-\frac{2801}{6804}\pi^4 \right)\nn \\
   & & +\left(\frac{m_{e}^2}{m_{\mu}^2}\right)^2 \left[\frac{4}{135}\log^5 \frac{m_{\mu}^2}{m_{e}^2}+\frac{17}{162}\log^4 \frac{m_{\mu}^2}{m_{e}^2}+\left(\frac{67}{81}+\frac{16}{81}\pi^2 \right)\log^3 \frac{m_{\mu}^2}{m_{e}^2}+\left(\frac{5237}{729}+\frac{34}{81}\pi^2 -\frac{32}{9}\zeta(3) \right)\log^2 \frac{m_{\mu}^2}{m_{e}^2}\right. \nn \\
    & & \hspace*{1.7cm} +\left(\frac{52153}{2187}+\frac{134}{81}\pi^2 +\frac{16}{45}\pi^4 -\frac{136}{27}\zeta(3)\right)\log \frac{m_{\mu}^2}{m_{e}^2}\nn \\ & & \left.\left. \hspace*{1.7cm}+\frac{1103423}{26244}+\frac{10474}{2187}\pi^2 +\frac{34}{135}\pi^4
-\frac{268}{27}\zeta(3)-\frac{64}{27}\pi^2 \zeta(3) -\frac{128}{9}\zeta(5)\right]
+\cO
 \left(\frac{m_{e}^2}{m_{\mu}^2}\right)^{5/2}\right\}
\\
& & \nn \\
& = &\left(\frac{\alpha}{\pi} \right)^5 20.142~813~2(5)\quad [{\it 20.142~93(23)}]\,.
\eea} } 

 %\vspace*{0.5cm}
{\footnotesize
{\setl
\bea
a_{\mu}^{(eee\mu)} & = & \left(\frac{\alpha}{\pi} \right)^5 \left\{ 
\left(\frac{119}{243}-\frac{4}{81}\pi^2 \right)\log^3 \frac{m_{\mu}^2}{m_{e}^2} -
\left(\frac{61}{243}-\frac{2}{81}\pi^2 \right)\log^2 \frac{m_{\mu}^2}{m_{e}^2}+
\left( \frac{7627}{1458}+\frac{52}{81}\pi^2 -\frac{16}{135}\pi^4\right)\log \frac{m_{\mu}^2}{m_{e}^2}\right. \nn \\ & &  \hspace*{1.cm} +\frac{64244}{6561}-\frac{2593}{2187}\pi^2 +\frac{8}{405}\pi^4 -\frac{476}{81}\zeta(3)+\frac{16}{27}\pi^2\zeta(3)\nn \\
& & + \left(\frac{m_{e}^2}{m_{\mu}^2}\right)\left[\left( -\frac{230}{81}+\frac{8}{27}\pi^2\right)\log^2 \frac{m_{\mu}^2}{m_{e}^2} +
\left(\frac{454}{27}-\frac{16}{9}\pi^2 \right)\log\frac{m_{\mu}^2}{m_{e}^2}-\frac{17525}{729}+\frac{76}{243}\pi^2
+\frac{32}{135}\pi^4 \right]\nn \\ 
& & +\left(\frac{m_{e}^2}{m_{\mu}^2}\right)^{3/2}\left(-\frac{41}{540}\pi^4\right)\nn \\ 
& & +\left(\frac{m_{e}^2}{m_{\mu}^2}\right)^2 \left[\frac{2}{45}\log^4 \frac{m_{\mu}^2}{m_{e}^2}+\left( \frac{2459}{2025}-\frac{4}{27}\pi^2 \right) \log^3 \frac{m_{\mu}^2}{m_{e}^2} +\left(\frac{19268}{3375}-\frac{26}{81}\pi^2 \right)\log^2 \frac{m_{\mu}^2}{m_{e}^2}\right. \nn \\ 
& & \hspace*{1.4cm} +\left(\frac{8929444}{455625}+\frac{98}{81}\pi^2 -\frac{16}{45}\pi^4 -\frac{32}{15}\zeta(3) \right)\log \frac{m_{\mu}^2}{m_{e}^2}\nn \\ & &  \hspace*{1.4cm}\left.\left. +
\frac{675818203}{13668750}-\frac{62}{243}\pi^2 -\frac{592}{2025}\pi^4 -\frac{3364}{225}\zeta(3) +\frac{16}{9}\pi^2 \zeta(3) \right] +\cO\left[\left(\frac{m_{e}^2}{m_{\mu}^2}\right)^{5/2}\right]\right\} \\
 & & \nn \\
  & = & \left(\frac{\alpha}{\pi} \right)^5 2.203~327~32(3)\quad [{\it 2.203~27(9)}]\,.
\eea}}

 %\vspace*{0.5cm}
{\footnotesize
{\setl
\bea
a_{\mu}^{(ee\mu\mu)} & = & \left(\frac{\alpha}{\pi} \right)^5 \left\{\left(-\frac{943}{486}-\frac{8}{405}\pi^2 +\frac{16}{9}\zeta(3) \right)\log^2 \frac{m_{\mu}^2}{m_{e}^2}+\left(\frac{57899}{7290}-\frac{10766}{6075}\pi^2 +\frac{8}{81}\pi^4 -\frac{8}{135}\zeta(3) \right)\log\frac{m_{\mu}^2}{m_{e}^2}\right.\nn \\ & & \hspace*{1.2cm} -\frac{1090561}{109350}-\frac{148921}{91125}\pi^2-\frac{106}{6075}\pi^4 +\frac{10732}{2025}\zeta(3)+\frac{32}{27}\pi^2 \zeta(3)\nn \\
 & & + \left(\frac{m_{e}^2}{m_{\mu}^2}\right)\left[\left(\frac{1832}{243}-\frac{104}{315}\pi^2 -\frac{32}{9}\zeta(3) \right)\log\frac{m_{\mu}^2}{m_{e}^2}-\frac{619798}{25515}+\frac{564008}{297675}\pi^2 -\frac{8}{81}\pi^4 +\frac{1328}{105}\zeta(3) \right] \nn \\
  & & 
 + \left(\frac{m_{e}^2}{m_{\mu}^2}\right)^2\left[\left(-\frac{29696}{45927}-\frac{852346}{535815}\pi^2 +\frac{8}{81}\pi^4 +\frac{3224}{567}\zeta(3)\right)\log\frac{m_{\mu}^2}{m_{e}^2}\right.\nn \\
& & \hspace*{1.6cm} \left. +\frac{51445307}{28934010}-\frac{546693856}{168781725}\pi^2 +\frac{26}{729}\pi^4 +\frac{752132}{178605}\zeta(3)+\frac{32}{27}\pi^2 \zeta(3)+\frac{64}{9}\zeta(5)  \right]\nn \\
& & \left.+\ \cO\left[\left(\frac{m_{e}^2}{m_{\mu}^2}\right)^3 \log^3 \frac{m_{\mu}^2}{m_{e}^2}\right]\right\}\\
& & \nn  \\
 & = & \left(\frac{\alpha}{\pi} \right)^5  0.206~959~089(2)\quad [{\it 0.206~97(2)}]
\eea}}

 %\vspace*{0.5cm}
{\footnotesize
{\setl
\bea
a_{\mu}^{(e\mu\mu\mu)} & = & \left(\frac{\alpha}{\pi} \right)^5\left\{ \left(\frac{151849}{30618}-\frac{8}{135}\pi^4 +\frac{128}{189}\zeta(3)  \right)\log\frac{m_{\mu}^2}{m_{e}^2}  \right. \nn \\ & & \hspace*{1.2cm} -\frac{46796257}{3214890}+\frac{143}{81}\pi^2 +\frac{124}{8505}\pi^4 +\frac{92476}{6615}\zeta(3)-\frac{16}{9}\pi^2 \zeta(3) \nn \\
 & & +\left(\frac{m_{e}^2}{m_{\mu}^2}\right)\left(-\frac{374711}{45927}-\frac{16}{675}\pi^2+\frac{16}{405}\pi^4 +\frac{2144}{567}\zeta(3) \right)\nn\\ 
 & & 
 +\left(\frac{m_{e}^2}{m_{\mu}^2}\right)^2 \left[\left(\frac{1565849}{5051970}+\frac{16}{525}\pi^2 -\frac{8}{405}\pi^4 +\frac{34064}{31185}\zeta(3) \right)\log\frac{m_{\mu}^2}{m_{e}^2} \right.  \nn \\
  & & \hspace*{1.6cm} \left. +\frac{16107486427}{70020304200}+\frac{5260603}{26790750}\pi^2 -\frac{11504}{467775}\pi^4 +\frac{652419088}{108056025}\zeta(3) -\frac{16}{27}\pi^2 \zeta(3)\right] \nn \\
  & & \left.+\ \cO\left[\left(\frac{m_{e}^2}{m_{\mu}^2}\right)^3 \log \frac{m_{\mu}^2}{m_{e}^2}\right]\right\}\\
& & \nn  \\
 & = & \left(\frac{\alpha}{\pi} \right)^5 0.013~875~909~09(6)\quad [{\it 0.013~88(1)}]\,.
\eea}}

 %\vspace*{0.5cm}
 
{\footnotesize
{\setl
\bea\lbl{Aeeetau}
a_{\mu}^{(eee\tau)} & = & \left(\frac{\alpha}{\pi} \right)^5 \left\{\left(\frac{m_\mu^2}{m_\tau^2}\right)\left[\frac{4}{1215}\log^3\frac{m_\mu^2}{m_{e}^2} -\frac{2}{405}\log^2 \frac{m_\mu^2}{m_{e}^2} -\left(\frac{122}{3645}-\frac{8\pi^2}{1215}\right)\log\frac{m_\mu^2}{m_{e}^2} \right.\right.  \nn \\
 & & \hspace*{2.2cm} \left. +\frac{2269}{32805}-\frac{4\pi^2}{215}-\frac{16}{405}\zeta(3)\right]   \nn \\
 & & +\left(\frac{m_e^2}{m_\tau^2} \right)\left[\frac{4}{45}\log^2 \frac{m_\mu^2}{m_{e}^2} -\frac{20}{27}\log\frac{m_\mu^2}{m_{e}^2} +\frac{634}{405}+\frac{8\pi^2}{135}\right]\nn\\  
 & & +\left(\frac{m_\mu^2}{m_\tau^2} \right)^2\left[-\frac{1}{945}\log^3\frac{m_\mu^2}{m_{e}^2} \log\frac{m_\tau^2}{m_\mu^2}-\frac{1}{630}\log^2\frac{m_\mu^2}{m_{e}^2}\log^2\frac{m_\tau^2}{m_\mu^2} +\frac{1}{945}\log\frac{m_\mu^2}{m_{e}^2} \log^3 \frac{m_\tau^2}{m_\mu^2}     \right. \nn \\
 & & \hspace*{1.6cm}-\frac{1}{3780}\log^4\frac{m_\tau^2}{m_\mu^2}+\frac{1}{14700}\log^3\frac{m_\mu^2}{m_{e}^2}-\frac{37}{33075}\log\frac{m_\tau^2}{m_e^2}\log\frac{m_\mu^2}{m_e^2}\log\frac{m_\tau^2}{m_\mu^2}  \nn \\
 & & \hspace*{1.6cm}-\frac{37}{99225}\log^3\frac{m_\tau^2}{m_\mu^2}+\left(\frac{229213}{1852200}+\frac{\pi^2}{945}\right)\log^2\frac{m_\mu^2}{m_e^2} \nn\\
 & & \hspace*{1.6cm} +\frac{40783}{3472875}\log\frac{m_\mu^2}{m_e^2}\log\frac{m_\tau^2}{m_\mu^2}+\frac{40783}{6945750}\log^2\frac{m_\tau^2}{m_\mu^2} \nn  \\
 & & \hspace*{1.6cm} -\left(\frac{30026659}{3889620000}-\frac{\pi^2}{945}\right)\log\frac{m_\mu^2}{m_e^2}+\frac{4}{315}\zeta(3)\log\frac{m_\tau^2}{m_\mu^2} \nn\\
& & \hspace*{1.6cm} \left. -\frac{24827672279}{544546800000} -\frac{59}{13608}\pi^2+\frac{4}{4725}\pi^4-\frac{1}{1225}\zeta(3)\right] \nn \\
 & & +\left(\frac{m_\mu^2}{m_\tau^2} \right)^3\left[-\frac{8}{8505}\log^3\frac{m_\mu^2}{m_{e}^2} \log\frac{m_\tau^2}{m_\mu^2}-\frac{4}{2835}\log^2\frac{m_\mu^2}{m_{e}^2}\log^2\frac{m_\tau^2}{m_\mu^2} +\frac{8}{8505}\log\frac{m_\mu^2}{m_{e}^2} \log^3 \frac{m_\tau^2}{m_\mu^2}     \right. \nn \\
 & & \hspace*{1.6cm}-\frac{2}{8505}\log^4\frac{m_\tau^2}{m_\mu^2}-\frac{524}{2679075}\log^3\frac{m_\mu^2}{m_{e}^2}-\frac{398}{893025}\log\frac{m_\tau^2}{m_e^2}\log\frac{m_\mu^2}{m_e^2}\log\frac{m_\tau^2}{m_\mu^2}  \nn \\
 & & \hspace*{1.6cm}-\frac{398}{2679075}\log^3\frac{m_\tau^2}{m_\mu^2}+\left(\frac{1102961}{112521150}+\frac{8\pi^2}{8505}\right)\log^2\frac{m_\mu^2}{m_e^2} \nn\\
 & & \hspace*{1.6cm} +\frac{2736946}{281302875}\log\frac{m_\mu^2}{m_e^2}\log\frac{m_\tau^2}{m_\mu^2}+\frac{1368473}{281302875}\log^2\frac{m_\tau^2}{m_\mu^2} \nn  \\
 & & \hspace*{1.6cm} -\left(\frac{311791591}{254441622500}+\frac{4\pi^2}{42525}\right)\log\frac{m_\mu^2}{m_e^2}+\frac{32}{2835}\zeta(3)\log\frac{m_\tau^2}{m_\mu^2} \nn\\
& & \hspace*{1.6cm} \left. -\frac{20302969165147}{446596444350000} -\frac{6299}{1913625}\pi^2+\frac{32}{42525}\pi^4+\frac{2096}{893025}\zeta(3)\right] \nn  \\ & & 
\left. +\ \cO\left[\left(\frac{m_{\mu}^2}{m_{\tau}^2}\right)^4\log^3 \frac{m_\mu^2}{m_\tau^2}\right]\right\} \\ 
 & = & \left(\frac{\alpha}{\pi} \right)^5 0.013~057~4(4)\,.
\eea}}

\end{appendix}

%\vspace*{1cm}

{\bf Acknowledgements}

\noi
The authors are grateful to Santi Peris for a careful reading of the manuscript.  The work of E.~de R. and D.~G. has been supported in part by the European Community's Marie Curie Research Training Network program under contract No. MRTN-CT-2006-035482, Flavianet. The work of D.~G. has also been supported by MEC (Spain) under grant FPA2007-60323, by Spanish Consolider-Ingenio 2010 Programme CPAN (CSD2007-00042) and he gratefully acknowledges an Experienced Researcher position supported by the EU-RTN Programme, contract No.MRTN--CT-2006-035482, Flavianet.

\end{document}